\documentclass[a4paper,11pt]{article}

\usepackage{jheppub}
\usepackage[T1]{fontenc}
\usepackage{amsmath,amssymb,mathtools}
\usepackage{booktabs}
\usepackage{array}
\usepackage{framed}
\usepackage{xcolor}
\usepackage{mdframed}
\usepackage{microtype}
\usepackage{amsthm}

\mdfdefinestyle{statementstyle}{
  frametitlerule=true,
  frametitlefont={\normalfont\bfseries},
  frametitlebackgroundcolor=blue!10,
}

\newcommand{\Lattice}{\Gamma}
\newcommand{\BPSCone}{\mathcal{K}_{\mathrm{BPS}}}
\newcommand{\Fixed}{\mathcal{F}}

\newcommand{\R}{\mathbb{R}}

\title{Reconstructing BPS String Charge Cone in \\ 6d $\mathcal{N}=(1,0)$ Supergravity}

\author[a,b,c]{Yuta Hamada}
\emailAdd{yhamada@post.kek.jp}
\author[d,e]{Hee-Cheol Kim}
\emailAdd{heecheol@postech.ac.kr}
\author[f]{Kai Xu}
\emailAdd{kaixu@tsinghua.edu.cn}

\affiliation[a]{Theory Center, IPNS, High Energy Accelerator Research Organization (KEK), 1-1 Oho, Tsukuba, Ibaraki 305-0801, Japan}
\affiliation[b]{Graduate Institute for Advanced Studies, SOKENDAI, 1-1 Oho, Tsukuba, Ibaraki 305-0801, Japan}
\affiliation[c]{RIKEN Center for Interdisciplinary Theoretical and Mathematical Sciences (iTHEMS), RIKEN, Wako 351-0198, Japan}
\affiliation[d]{Department of Physics, POSTECH, Pohang 37673, Korea}
\affiliation[e]{Asia Pacific Center for Theoretical Physics, Postech, Pohang 37673, Korea}
\affiliation[f]{Yau Mathematical Science Center, Tsinghua University, Beijing 100084, China}

\subheader{%
  \makebox[\textwidth][r]{%
    \normalfont\small
    \begin{tabular}{@{}r@{}}
      KEK-TH-2876\\
      RIKEN-iTHEMS-Report-26
    \end{tabular}%
  }%
}

\abstract{The tensor branch of six-dimensional $\mathcal N=(1,0)$ supergravity is characterized by a spectrum of BPS strings charged under the two-form gauge fields. Their charges span a cone in the tensor charge lattice, whose extremal rays are generated by primitive BPS string charges that become tensionless at the boundaries of the tensor moduli space. We develop a systematic procedure, based entirely on effective field theory data, to reconstruct the primitive BPS string generators and thereby determine the complete closed BPS string cone of any 6d $(1,0)$ supergravity theory. The key ingredients are the universal {\it H}-string, which allows us to isolate a fixed sector determined by classified LST blocks and gauge anomaly data from the remaining gauge-neutral E-string sector, and a BPS refinement of the Completeness Conjecture that ensures population of the relevant charge rays. We further show that this reconstruction provides new constraints on the low-energy theory and can rule out candidate EFTs that pass all previously known local lattice and anomaly consistency conditions. Separately, for theories admitting geometric tensor bases, we explain how the same reconstruction determines the complete Mori cone of the corresponding K\"ahler surface.}

\begin{document}
\hypersetup{pageanchor=false}
\maketitle

\hypersetup{pageanchor=true}
\raggedbottom

\section{Introduction}
\label{sec:introduction}

Six-dimensional supergravity theories with eight supercharges provide an exceptionally rigid framework for exploring the consistency conditions of quantum gravity. Their effective descriptions are tightly constrained by anomaly cancellation, supersymmetry, charge quantization, and unitarity, while at the same time exhibiting intrinsically non-perturbative phenomena such as tensionless strings and strongly coupled fixed points.  These features make six-dimensional $\mathcal N=(1,0)$ supergravity an especially sharp setting for the Swampland program, whose aim is to distinguish effective field theories that admit a consistent ultraviolet completion from those that do not~\cite{Vafa:2005ui,Agmon:2022thq}.

A central lesson from recent developments is that perturbative consistency conditions alone are not sufficient to determine the structure of a consistent six-dimensional theory of quantum gravity  \cite{Kim:2019vuc,Kim:2024eoa}.  Crucial additional information is supplied by BPS strings that become tensionless at boundaries of the tensor moduli space, where the effective field theory description breaks down.  Finite-distance boundaries are associated with local 6d superconformal field theories (SCFTs), while infinite-distance limits can be governed by little string theories (LSTs) or critical strings~\cite{Heckman:2015bfa,Bhardwaj:2015xxa,Bhardwaj:2015oru,Bhardwaj:2019hhd,Lee:2019xtm}.  The spectrum and intersections of these strings therefore encode non-perturbative information about the UV completion, beyond what is captured by the massless spectrum.

This insight was promoted to a concrete organizing principle in~\cite{Kim:2024eoa}, based on the assumption that every boundary of the tensor moduli space of a consistent 6d (1,0) supergravity theory supports BPS strings that become tensionless. The charges of physical BPS strings span a cone, the BPS cone, dual to the tensor cone, whose extremal rays are generated by primitive BPS strings that can become tensionless somewhere in the tensor moduli space. Together with the classifications of 6d SCFTs and LSTs~\cite{Heckman:2015bfa,Bhardwaj:2015xxa,Bhardwaj:2015oru,Bhardwaj:2019hhd,Lee:2019xtm}, this principle strongly constrains the tensor charge lattice and its intersection structure. This framework has already led to powerful results on the boundedness of the tensor sector and the finiteness of the massless spectrum~\cite{Kim:2024eoa,Hamada:2026zta,Kim:2026ldd}. These developments further motivate the prospect that the full tensor moduli space should be recoverable from the spectrum of primitive BPS strings using only low-energy EFT information.

The aim of this work is to realize this perspective. We develop a bottom-up framework for reconstructing the BPS string charge cone and, consequently, the full tensor moduli space using only six-dimensional EFT data without assuming any underlying string theory or geometric realization. The central organizing object is the \textit{H}-string introduced in~\cite{Kim:2024eoa}, which was shown to exist in every six-dimensional (1,0) supergravity with $T\geq1$ and $b_0\neq0$, where $b_0$ denotes the gravitational anomaly charge. This string has a null charge $f$ satisfying $f^2=0$ and $b_0\cdot f=2$.\footnote{Strings with these charge properties were previously
used in~\cite{Lee:2019skh} to constrain the number of $U(1)$
gauge factors in 6d supergravity.} It becomes tensionless at an infinite-distance boundary, and its worldsheet anomaly data agree with those of a critical heterotic string compactified to six dimensions or a little string.

The existence of the {\it H}-string reduces the reconstruction problem to a much more tractable form.
It naturally divides the negative-norm primitive BPS spectrum into an $f$-orthogonal sector and a sector with positive $f$-intersection. The generators in the first sector belong to the LSTs associated with the \textit{H}-string and can therefore be determined from EFT data using the LST classification. Within the second sector, generators supporting non-trivial gauge algebras are likewise fixed by the corresponding gauge anomaly coefficients and worldsheet current algebras. Together, these two classes form a {\it fixed set} of primitive generators whose charges and mutual intersections are already determined from intrinsic EFT data.  The only generators not determined in this way have positive intersection with the \textit{H}-string but support no gauge algebra. EFT consistency conditions and the classification of 6d SCFTs constrain such generators to be E-strings, or equivalently the $(-1)$-generators without gauge algebra studied, for example, in ~\cite{Ganor:1996mu,Minahan:1998vr,Kim:2014dza}. The remaining task is to determine which candidate E-string charges are populated and span primitive extremal rays of the BPS cone.

The E-string sector is thus the only residual ambiguity in the reconstruction, and resolving it completes the determination of the full BPS string charge cone. Once the fixed generators are specified, candidate E-string charges can be systematically enumerated by imposing the characteristic numerical conditions of an E-string together with non-negative intersection with every fixed generator. In general, however, these conditions admit too many candidate charges, all of which appear equally admissible from the viewpoint of the charge lattice and the perturbative EFT constraints. The essential problem is therefore not simply to enumerate E-string candidates, but to determine which collections of them are mutually compatible and can simultaneously occur as primitive generators of a physical BPS cone.

The {\it H}-string provides the additional structure needed to organize this ambiguity. Its intersection with an E-string candidate defines a natural degree, which in turn orders competing candidates. We show that the numerical candidates organize into maximal compatible sets and that there always exists a unique {\it lowest-ordered set}, $\mathcal S_{\rm low}$, selected by this $H$-string ordering. The existence and uniqueness of this set are properties of the reconstruction itself and do not assume that its charges are already populated. The remaining physical question is whether the candidate charges selected in this way are actually occupied by dynamical BPS strings.

To determine whether the candidates in this distinguished set are physically realized, we introduce the \emph{BPS Ray Completeness Hypothesis}, a BPS refinement of the Completeness Hypothesis~\cite{Polchinski:2003bq,Banks:2010zn,Harlow:2018tng}. It asserts that every charge direction that is locally admissible as a supersymmetric string source in the effective theory contains a non-zero integral charge realized by a dynamical BPS configuration. Provided that local supersymmetric source admissibility is established for every candidate in the lowest-ordered set, BPS Ray Completeness ensures that a positive integral multiple of each candidate is BPS-populated. Together with the complete classification of tensionless sectors and their BPS string spectra, this promotes the lowest-ordered candidates to primitive BPS generators. Consequently, the lowest-ordered E-string set together with the fixed generators determines the complete closed BPS string charge cone. The details of the candidate construction, compatible sets, $H$-string ordering, and population argument are developed in Sections~\ref{sec:E-string-set} and~\ref{sec:eft}.

Our main result is therefore that a small set of intrinsic physical criteria uniquely determines all BPS generators and reconstructs the BPS string charge cone. The required EFT data consist of the tensor charge lattice, the gravitational and gauge anomaly coefficients, and the fixed sectors associated with the {\it H}-string LSTs and gauge-supporting tensor multiplets. All of these are determined entirely by low-energy EFT data. In the general case, the reconstruction is supplemented by the BPS Ray Completeness Hypothesis. No underlying string-theoretic or geometric realization is assumed. The construction is therefore entirely bottom-up within six-dimensional effective field theory.

We provide several pieces of evidence for BPS Ray Completeness from the EFT perspective. On an auxiliary rational K\"ahler surface realizing the fixed tensor data, Riemann--Roch makes every numerical E-string candidate effective, while the $H$-string ordering identifies all external E-string Mori generators. This will show that whenever the tensor sector admits the rational-surface realization described in Section~\ref{sec:geometry}, the same reconstruction reproduces the complete Mori cone of the surface. Effectivity on an auxiliary surface alone, however, does not establish BPS population in the original EFT. For an F-theory base, this geometric cone is identified with the physical BPS string cone through wrapped D3-branes.

We also illustrate the reconstruction in several explicit examples, including both geometrically realized and apparently non-geometric tensor sectors. In particular, we exhibit a seemingly consistent 6d EFT satisfying the previously known perturbative and non-perturbative consistency conditions that is nevertheless excluded once the complete BPS cone is taken into account. This demonstrates that the existence of a consistent BPS string charge cone supplies an additional non-perturbative Swampland constraint on 6d $\mathcal N=(1,0)$ supergravity.

The paper is organized as follows. In Section~\ref{sec:physical-setup}, we review the tensor charge lattice, the fixed $H$-string LST and gauge sectors, and the physical assumptions used in the reconstruction. Section~\ref{sec:E-string-set} constructs the numerical E-string candidates and their compatible sets. Section~\ref{sec:eft} proves the existence of the unique lowest set, explains causal completion, derives the general EFT uniqueness theorem, and establishes the independent low-$T$ and null-boundary criteria. Section~\ref{sec:reconstruction} gives the reconstruction algorithm and examples. Section~\ref{sec:geometry} collects the separate geometric interpretation and realization problem. We conclude in Section~\ref{sec:conclusion}.

\section{Six-dimensional supergravity and BPS strings}
\label{sec:physical-setup}

In this section we review the basic ingredients that underlie our reconstruction of the BPS string charge cone in 6d $\mathcal{N}=(1,0)$ supergravity. We first summarize the massless field content, anomaly cancellation, and tensor charge lattice, and then review the tensor moduli space and the role of BPS strings. Finally, we explain the refined tensor intersection structure associated with the {\it H}-string and introduce the notion of the fixed set of primitive generators that will serve as the starting point of our reconstruction.

\subsection{Massless spectrum, anomaly cancellation, and anomaly vectors}
\label{sub:lattice-anomaly-data}

The massless spectrum of a six-dimensional $\mathcal N=(1,0)$ supergravity theory consists of a gravity multiplet together with tensor, vector, and hypermultiplets. The gravity multiplet contains the graviton and a self-dual two-form tensor field, while each tensor multiplet contains an anti-self-dual two-form tensor together with a real scalar. Vector multiplets furnish the gauge algebra $\mathfrak{g}=\oplus_i\,\mathfrak{g}_i$, and hypermultiplets provide the charged and neutral matter fields.

Among the fields of six-dimensional supergravity, tensor multiplets play a particularly important role in our analysis. Their scalar fields span the tensor branch of the moduli space, and their associated two-form gauge fields couple to strings. Consequently, each tensor multiplet contributes both a scalar modulus and the corresponding string charges. The tensor sector thus governs both the structure of the tensor moduli space and the lattice of BPS string charges. The string charges take values in a Lorentzian lattice $\Gamma$ of signature $(1,T)$. Dirac quantization imposes integrality of the lattice pairing, whereas consistency of the chiral two-form fields upon compactification requires the charge lattice to be self-dual~\cite{Seiberg:2011dr}, namely, $\Gamma=\Gamma^*$.

The massless matter content is subject to strong consistency conditions arising from the cancellation of local gauge, gravitational, and mixed anomalies. These anomalies are cancelled through the Green-Schwarz-Sagnotti mechanism, in which the one-loop anomaly polynomial factorizes into a quadratic form constructed from the two-form gauge fields~\cite{Green:1984sg,Sagnotti:1992qw,Erler:1993zy} as follows:
\begin{equation}
 I_8=\frac12\Omega_{\alpha\beta}X_4^\alpha X_4^\beta \ ,
 \qquad
 X_4^\alpha=-\frac12b_0^\alpha\operatorname{tr}R^2
 +\frac14\sum_i b_i^\alpha\frac{2}{\lambda_i}\operatorname{tr}F_i^2\,.
 \label{eq:anomaly-factorization}
\end{equation}
Here, $\Omega$ denotes the tensor intersection form, $b_0$ is the gravitational anomaly vector, each $b_i$ is associated with a gauge factor $G_i$ with field strength $F_i$, and $\lambda_i$ fixes the standard normalization of the gauge trace.

The factorization conditions determine the inner products among the anomaly vectors entirely in terms of the massless spectrum. In particular, cancellation of the gravitational anomaly requires~\cite{Erler:1993zy,Kumar:2010ru}
\begin{equation}
    H-V+29T=273\ , \qquad b_0^2=9-T\,,
 \label{eq:b0-square}
\end{equation}
where $T, V$, and $H$ are the numbers of tensor, vector, and hypermultiplets, respectively, and $b_0^2 \equiv \Omega_{\alpha\beta}b_0^\alpha b_0^\beta$.  Similarly, gauge and mixed anomalies determine $b_i^2$, $b_0\cdot b_i$, and $b_i\cdot b_j$. Note that the norm of $b_0$ is fixed by the number of tensor multiplets, while the inner products involving the gauge anomaly vectors are determined by the representations and multiplicities of charged hypermultiplets. In particular, when two tensor charges support gauge algebras, they are identified with the corresponding anomaly vectors $b_i$ and $b_j$, and their mutual intersection is completely fixed by the spectrum of bifundamental matter through the mixed gauge anomaly. Thus, anomaly cancellation alone already determines a substantial part of the tensor intersection structure.

The anomaly vectors span an integral sublattice of the full string charge lattice, commonly referred to as the anomaly lattice~\cite{Kumar:2010ru}. Remarkably, although the anomaly equations are expressed in terms of rational group-theoretic coefficients, the resulting inner products are always integral~\cite{Monnier:2017oqd}. Thus, the Gram matrix of the anomaly vectors must possess at most one positive eigenvalue and no more than $T$ negative eigenvalues~\cite{Hamada:2023zol}. These conditions provide highly non-trivial constraints on the admissible matter spectrum and the tensor sector of every consistent 6d supergravity.

\subsection{Tensor moduli space and BPS cone}

The expectation values of the tensor multiplet scalars are collectively described by a vector $J\in\R^{1,T}$ satisfying $J^0>0$ and $J^2=1$. Locally, these scalars parameterize the hyperbolic space $SO(1,T)/SO(T)$, but only a proper subregion corresponds to physical vacua~\cite{Romans:1986er,Seiberg:1996vs}. This region is determined by positivity of BPS string tensions. A string with tensor charge $Q\in\Lattice$ has tension proportional to $J\cdot Q$. We call a lattice charge $Q$ \emph{populated} if it is carried by a dynamical BPS configuration: either a single string or a finite aligned multi-string configuration preserving the same supersymmetry. This does not require a single-string bound state~\cite{Haghighat:2015ega}. This intrinsic EFT notion should be distinguished from \emph{effectivity}, which means that $Q$ is represented by an effective curve in a geometric realization. Hence every point in the physical tensor moduli space must obey
\begin{equation}\label{eq:cone}
    J\cdot Q\ge0
\end{equation}
for every populated BPS string charge. The spectrum of BPS strings therefore determines the physical region of the tensor moduli space.

Following the proposal of~\cite{Kim:2024eoa}, we assume throughout this paper that every boundary of the tensor moduli space supports BPS strings that become tensionless there. Finite-distance boundaries are associated with local 6d SCFTs, whereas infinite-distance boundaries are governed by little string theories or critical strings. Under this assumption, every boundary face of the tensor moduli space at which the effective supergravity description breaks down is supported by at least one primitive BPS string whose tension vanishes there. Consequently, the global structure of the tensor moduli space is completely determined by the primitive BPS string spectrum.

The charges of physical BPS strings generate a closed convex cone in the real tensor charge space, which we call the \emph{BPS cone}. Throughout this paper, $\operatorname{Cone}(A)$ denotes the closed conical hull:
\[
 \operatorname{Cone}(A)
 :=
 \overline{\left\{
 \sum_{i=1}^{N}a_iQ_i
 \,\middle|\,
 N<\infty,\ a_i\geq0,\ Q_i\in A
 \right\}}.
\]
We use the physical generator presentation in which this closed cone is generated by the primitive extremal BPS string charges $C_i$, called \emph{BPS generators}. Thus
\begin{equation}
 \BPSCone
 =\operatorname{Cone}\{Q\in\Lattice\mid Q\text{ is BPS-populated}\}
 =\operatorname{Cone}\{C_i\}.
 \label{eq:bps-cone-generators}
\end{equation}
The closure includes accumulation directions when the generator set is infinite. We note that membership in $\BPSCone$ does not by itself imply that a charge, or any positive integral multiple of it, is BPS-populated. We also use the physical dual-cone condition
\begin{equation}
 \mathcal T=\BPSCone^\vee\subseteq\overline{\mathcal C^+}\,,
 \label{eq:dual-cone-causality}
\end{equation}
where $\overline{\mathcal C^+}$ denotes the closed future Lorentz cone. The cone $\mathcal T$ is the tensor cone, and the physical tensor moduli space is its unit hyperbolic slice defined by $J^2=1$. Physically, this states that every tensor direction compatible with non-negative BPS string tensions is future causal. In particular, the dual tensor cone contains no spacelike direction.

Once all BPS generators have been identified, positivity of the tension of every BPS string is equivalent to positivity for all these generators. The physical tensor moduli space is thus given by
\begin{equation}
 \mathcal{M}_T
 =\bigl\{J\,\big|\,J^0>0,\ J^2=1,\ J\cdot C_i\geq0
 \text{ for every BPS generator }C_i\bigr\} \, .
 \label{eq:tensor-moduli}
\end{equation}
Under the physical boundary assumptions above, each BPS generator defines either a finite-distance tensionless wall, where $J\cdot C_i=0$ when $C^2_i<0$, or an asymptotic tensionless limit, where $J\cdot C_i\rightarrow 0$ when $C_i^2=0$. It follows that every BPS generator has non-positive norm
\begin{equation}
 C_i^2\leq0 \, .
 \label{eq:generator-nonpositive-norm}
\end{equation}
Moreover, distinct BPS generators must have non-negative mutual intersections~\cite{Kim:2024eoa}:
\begin{equation}
C_i\cdot C_j\geq0\,,\qquad \text{for }i\neq j\, .
\label{prop:mutual-nonnegativity}
\end{equation}

The BPS generators are exhausted by the known strings arising in local SCFTs for $C_i^2<0$, and in LSTs or critical-string theories for $C_i^2=0$, all of which have been classified, for example,  in~\cite{Heckman:2015bfa,Bhardwaj:2015xxa,Bhardwaj:2015oru,Bhardwaj:2019hhd}. These generators comprise instanton strings with charges $C_i=b_i$ supporting non-trivial gauge algebras, together with four gauge-free classes: the E-string, the M-string, the critical heterotic string, and the Type II string. Table~\ref{tab:bps-generator-classification} presents the complete classification. The first four rows correspond to the gauge-free generators, while the remaining rows list the instantonic generators associated with gauge algebras $\mathfrak{g}_i$.

We assume that the classification of 6d SCFTs and LSTs is complete, including the BPS string charge spectra of their tensionless sectors. This assumption concerns not only primitive generators but also the allowed multiply charged and composite BPS string sectors. Thus a tensionless configuration must belong to a classified local theory with its associated string spectrum, rather than merely satisfy the numerical conditions of a classified primitive charge.

Among the generators in this list, the M-string requires special treatment, as discussed in~\cite{Kim:2024eoa,Hamada:2026zta}. When an M-string charge $C$ becomes tensionless, a Weyl reflection identifies the two neighboring tensor chambers. One may therefore extend the marked tensor moduli space through this wall and remove the associated M-string charge from the primitive generator set. This removal is not allowed, however, when the $\mathfrak{su}_2$ flavor symmetry of the M-string is gauged, as occurs in the $(-2,-3), (-2,-2,-3)$, and $(-2,-3,-2)$ non-Higgsable chains. Moreover, if a charge satisfying $C^2=-2$ and $b_0\cdot C=0$ supports a gauge algebra, the corresponding string is not an M-string but an instanton string of that gauge algebra. Such strings remain physical BPS generators.

\begin{table}[t]
\centering
\scriptsize
\setlength{\tabcolsep}{3pt}
\renewcommand{\arraystretch}{1.08}
\begin{tabular}{@{}l>{\raggedright\arraybackslash}p{0.47\textwidth}ccc@{}}
\toprule
String or $\mathfrak g_i$ & $H_i$ & $C_i^2$ & $b_0\cdot C_i$ & Restriction \\
\midrule
E-string & --- & $-1$ & $1$ &  \\
M-string & --- & $-2$ & $0$ &  \\
heterotic string & --- & $0$ & $2$ & \\
Type II string & --- & $0$ & $0$ &  \\
\midrule
$\mathfrak g$ & ${\bf Adj}$ & $0$ & $0$ & \\
$\mathfrak{su}_3$ & --- & $-3$ & $-1$ & \\
$\mathfrak{su}_N$ & $2N\times\mathbf N$ & $-2$ & $0$ & \\
$\mathfrak{su}_N$ & $(N-8)\times\mathbf N\oplus\mathbf{N(N+1)}/2$ & $-1$ & $-1$ & $N\geq8$ \\
$\mathfrak{su}_N$ & $(N+8)\times\mathbf N\oplus\mathbf{N(N-1)}/2$ & $-1$ & $1$ & \\
$\mathfrak{su}_N$ & $16\times\mathbf N\oplus2\times\mathbf{N(N-1)}/2$ & $0$ & $2$ & \\
$\mathfrak{su}_N$ & $\mathbf{N(N-1)}/2\oplus\mathbf{N(N+1)}/2$ & $0$ & $0$ & \\
$\mathfrak{su}_6$ & $15\times\mathbf6\oplus\tfrac12\mathbf{20}$ & $-1$ & $1$ & \\
$\mathfrak{su}_6$ & $17\times\mathbf6\oplus\mathbf{15}\oplus\tfrac12\mathbf{20}$ & $0$ & $2$ & \\
$\mathfrak{su}_6$ & $18\times\mathbf6\oplus\mathbf{20}$ & $0$ & $2$ & \\
$\mathfrak{su}_6$ & $\mathbf6\oplus\tfrac12\mathbf{20}\oplus\mathbf{21}$ & $0$ & $0$ & \\
$\mathfrak{so}_N$ & $(N-8)\times\mathbf N$ & $-4$ & $-2$ & $N\geq8$ \\
$\mathfrak{so}_N$ & $(N-7)\times\mathbf N\oplus2^{\lfloor(10-N)/2\rfloor}\times\mathbf{2^{\lfloor(N-1)/2\rfloor}}$ & $-3$ & $-1$ & $7\leq N\leq12$ \\
$\mathfrak{so}_N$ & $(N-6)\times\mathbf N\oplus\bigl(2\times2^{\lfloor(10-N)/2\rfloor}\bigr)\times\mathbf{2^{\lfloor(N-1)/2\rfloor}}$ & $-2$ & $0$ & $6\leq N\leq13$ \\
$\mathfrak{so}_N$ & $(N-5)\times\mathbf N\oplus\bigl(3\times2^{\lfloor(10-N)/2\rfloor}\bigr)\times\mathbf{2^{\lfloor(N-1)/2\rfloor}}$ & $-1$ & $1$ & $5\leq N\leq12$ \\
$\mathfrak{so}_N$ & $(N-4)\times\mathbf N\oplus\bigl(4\times2^{\lfloor(10-N)/2\rfloor}\bigr)\times\mathbf{2^{\lfloor(N-1)/2\rfloor}}$ & $0$ & $2$ & $4\leq N\leq14$ \\
$\mathfrak{sp}_N$ & $(2N+8)\times\mathbf{2N}$ & $-1$ & $1$ & \\
$\mathfrak{sp}_N$ & $16\times\mathbf{2N}\oplus\mathbf{(N-1)(2N+1)}$ & $0$ & $2$ & \\
$\mathfrak{sp}_3$ & $\tfrac{35}{2}\mathbf6\oplus\tfrac12\mathbf{14}'$ & $0$ & $2$ & \\
$\mathfrak e_8$ & --- & $-12$ & $-10$ & \\
$\mathfrak e_7$ & $\tfrac{k}{2}\times\mathbf{56}$ & $k-8$ & $k-6$ & $k\leq8$ \\
$\mathfrak e_6$ & $k\times\mathbf{27}$ & $k-6$ & $k-4$ & $k\leq6$ \\
$\mathfrak f_4$ & $k\times\mathbf{26}$ & $k-5$ & $k-3$ & $k\leq5$ \\
$\mathfrak g_2$ & $(3k+1)\times\mathbf7$ & $k-3$ & $k-1$ & $k\leq3$ \\
\bottomrule
\end{tabular}
\caption{Classification of BPS generators. The first four rows are gauge-free generators, while the remaining charges correspond to instanton strings associated with gauge algebras $\mathfrak{g}_i$. The charged hypermultiplet content is denoted by $H_i$.  Bold symbols denote representations by their dimensions.}
\label{tab:bps-generator-classification}
\end{table}

The gravitational anomaly charge $b_0$ provides another distinguished element of the tensor charge lattice, even though it may not be primitive and is not itself necessarily a BPS generator. The supersymmetric-defect argument of~\cite{Kim:2024eoa} implies that $n b_0$ is populated for some positive integer $n$. Since $\BPSCone$ is a real cone, it follows that
\begin{equation}
 b_0\in\BPSCone\,.
 \label{eq:b0-cone-membership}
\end{equation}
This does not imply that the charge $b_0$ itself is populated or defines a BPS generator. Consequently, $b_0$ satisfies
\begin{equation}
 J\cdot b_0\geq0
 \label{eq:b0-effective}
\end{equation}
throughout the tensor moduli space.
See also~\cite{Cheung:2016wjt,Hamada:2018dde} for related
positivity arguments involving the Gauss--Bonnet coupling
and black-hole entropy.

\subsection{Refined tensor intersection structure}
\label{sub:refined-tensor-structure}

A major advance in understanding the tensor sector of 6d supergravity was the discovery of a distinguished BPS string, called the {\it H}-string, together with a refined structure of the tensor moduli space developed in~\cite{Kim:2024eoa,Hamada:2026zta,Kim:2026ldd}. These results reveal a universal organization of the BPS generators that goes substantially beyond perturbative considerations such as anomaly cancellation. This structure serves as the starting point for our reconstruction framework.

The {\it H}-string exists in every 6d $(1,0)$ supergravity with $T\ge1$, except for the special case $T=9$ with $b_0=0$ \cite{Kim:2024eoa}. Its charge $f$ is characterized by
\begin{equation}
 f^2=0\,,
 \qquad
 b_0\cdot f=2\,.
 \label{eq:h-string}
\end{equation}
The worldsheet anomalies can be determined from the inflow analysis of~\cite{Kim:2016foj,Shimizu:2016lbw,Kim:2019vuc}, and one finds the worldsheet central charges $c_L=20, c_R=6$ of the worldsheet CFT on the {\it H}-string. These values precisely match those of a critical heterotic string compactified to six dimensions or a fundamental little string. 

The \textit{H}-string becomes tensionless in the asymptotic region $J\sim f$ of the tensor moduli space. For $T=1$, the charge $f$ can itself generate an extremal ray of the BPS cone, whereas for $T>1$ it is the total charge of a little string and can be decomposed into primitive BPS string charges contained in the associated little string theory.

The existence of the {\it H}-string follows from a sequence of physical blowdowns.  Let $C$ be a primitive $(-1)$-generator with $C^2=-1$ and $b_0\cdot C=1$. Restricting to the tensionless wall associated with $C$, i.e., $J\cdot C=0$, projects the charge lattice onto the subspace orthogonal to $C$. Under this projection, every charge transforms as
\begin{equation}
 Q\longmapsto Q'=Q+(C\cdot Q)C\, , \qquad Q'\cdot C=0\, .
\label{eq:eft-blowdown}
\end{equation}
This operation removes one tensor direction, with the anomaly vectors $b_0, b_i$ transforming according to the same rule.  If the reduced BPS cone contains another $(-1)$-generator, the blowdown can be repeated. The sequence terminates either at a two-dimensional tensor cone or a cone with no further  blowdown. Combined with the real BPS-cone condition~\eqref{eq:b0-effective} for the gravitational anomaly vector and the classification of BPS generators in Table~\ref{tab:bps-generator-classification}, this analysis requires the existence of an \textit{H}-string charge satisfying~\eqref{eq:h-string}.

Moreover, the containment relation established in~\cite{Kim:2024eoa} tells us that the \textit{H}-string organizes the complete set of BPS generators according to their intersection with its charge $f$. First, every BPS generator must have non-negative intersection with $f$. Then, any generator orthogonal to the {\it H}-string necessarily belongs to one of the LSTs carrying the common little string charge $f$. Conversely, within each such LST sector, the \textit{H}-string charge decomposes as
\begin{equation}
 f=\sum_i n_i C_i\, ,
 \qquad C_i\cdot f=0\, ,
 \qquad n_i\in\mathbb Z_{>0}\,,
 \label{eq:h-string-lst-decomposition}
\end{equation}
where $C_i$ denotes the collection of BPS string charge generators in the LST. All remaining generators lie outside these LST sectors and must have strictly positive intersection with $f$. We will refer to such generators with $f\cdot C>0$ as ``{\it external generators}''. The tensor intersection structure therefore separates into an LST sector, determined entirely by the known classification of LSTs, and an external sector consisting of generators with positive intersection with $f$. This decomposition is completely intrinsic to the tensor charge lattice of 6d supergravity and does not rely on any geometric realization.

Taken together, these results prove the following universal statement, except for theories with $T=9$ and $b_0=0$~\cite{Kim:2024eoa}:
\begin{framed}
    \centering
    {\it For every six-dimensional supergravity theory, the tensor intersection structure \\
    is that of K\"ahler surfaces: $\mathbb{P}^2$, $\mathbb{F}_n$, or their blowups.}
\end{framed}
The blowdown analysis above makes this correspondence explicit. In particular, the topology of the tensor charge lattice encoded in $(\Omega,b_0,b_i)$ coincides with that of these K\"ahler surfaces. For $T=0$, the tensor intersection data are those of $\mathbb P^2$, while for $T=1$, they are those of a Hirzebruch surface $\mathbb F_n$. For $T>1$, reversing the blowdowns reconstructs the intersection structure of a blowup of one of these surfaces. Importantly, this conclusion follows entirely from effective field theory and assumes no underlying geometric realization. In particular, it does not imply that the complete BPS cone of every supergravity theory admits a geometric description. 

So far, no general procedure has been available for reconstructing the full BPS cone directly from EFT data. Moreover, the lattice pairing and the anomaly data $(\Omega,b_0,b_i)$ may not by themselves be sufficient to identify the BPS cone of a 6d EFT with the Mori cone of a particular surface. Constructing the full BPS cone from EFT data and proving this stronger geometric identification are the central goals of the following sections.

\subsection{Fixed generators and remaining E-string sector}
\label{sub:fixed_generators}

The refined tensor intersection structure reviewed in the previous subsection considerably reduces the amount of information that must be reconstructed. Rather than determining the entire BPS spectrum at once, which would be extremely challenging, one can first identify a distinguished subset of BPS generators that can be systematically fixed by universal EFT data explained above. This subset provides the starting point of our reconstruction of the BPS cone and will be referred to as the {\it fixed set} of generators.

The first contribution to the fixed set comes from the LST sector associated with the {\it H}-string. As discussed above, every primitive generator orthogonal to the {\it H}-string belongs to one of the LSTs sharing the common little string charge $f$. We denote the set of these generators by
\begin{equation}
 \Fixed_{\mathrm{LST}}(f)
 :=\{C_i\mid C_i\cdot f=0\},
 \label{eq:fixed-lst}
\end{equation}
where $C_i$ runs over the primitive BPS generators. Since these LSTs are completely classified, this set and all mutual intersections among its generators are already determined. No additional reconstruction is required for this sector.

The second contribution consists of {\it external} BPS generators, which satisfy $f\cdot C_i>0$ and support non-Abelian gauge algebras. Their charges are identified with the anomaly vectors associated with the corresponding tensor multiplets, and their mutual intersections are fixed by gauge anomaly cancellation through the spectrum of bifundamental matter. Consequently, once the gauge sector of a 6d theory is specified, these generators and their intersection numbers are likewise determined entirely from low-energy EFT data. We also include in this class the external M-strings whose $\mathfrak{su}_2$ flavor symmetry is gauged. They arise only in the $(-2,-3)$, $(-2,-2,-3)$, and $(-2,-3,-2)$ NHC chains~\cite{Morrison:2012np}, so their intersections with the other generators are likewise fixed by EFT data. We denote this combined set of external generators by $\Fixed^{+}(f)$. No external E-string is placed in the initial fixed set. These E-string charges are precisely the sector reconstructed below.

We define the ``\textit{fixed set of generators}'' as the union of these two sectors:
\begin{equation}
 \Fixed=\Fixed_{\mathrm{LST}}(f)\cup\Fixed^{+}(f)
 =\{F_a\}.
 \label{eq:fixed-set}
\end{equation}
Every element of \(\Fixed\) is determined directly from intrinsic EFT information, namely the tensor charge lattice, the anomaly coefficients, and the classifications of six-dimensional SCFTs and LSTs. Furthermore, all mutual intersections within the fixed set are known before reconstructing the remaining BPS sector. We also note that the intersection numbers between the external generators in the second class $\Fixed^{+}$ and E-strings in the LST sectors are proven to be bounded in~\cite{Kim:2026ldd}. 

The fixed set must obey the universal intersection rules reviewed above. For any proposed fixed set, we must also verify that its intersection structure admits the physical blowdown sequence to a Hirzebruch endpoint $\mathbb F_b$ described above. These local and topological tests are necessary, but they do not establish consistency of the full tensor sector. One must still reconstruct the complete BPS cone and check its global properties, including membership of the gravitational charge $b_0$. The proposed $T=11$ configuration in Section~\ref{sub:T11-global-obstruction} illustrates the importance of this additional step: it passes the preceding tests, but its reconstructed cone fails to contain the gravitational charge $b_0$.

To carry out this global test, we must therefore determine which BPS generators complete $\Fixed$ to the full BPS cone. The remaining sector is sharply constrained. Since $\Fixed$ already contains every generator orthogonal to $f$ and every external generator supporting a gauge algebra, any additional generator must have $f\cdot C>0$ and carry no gauge algebra. M-strings with ungauged $\mathfrak{su}_2$ flavor symmetry have already been removed by Weyl reflection. The classification in Table~\ref{tab:bps-generator-classification} therefore leaves only external E-strings satisfying the worldsheet consistency conditions. Identifying which of their charges are physical generators is the remaining reconstruction problem, and supplies the missing information needed to test the complete tensor sector.

For $T>1$, the independent primitive generators that remain to be reconstructed have negative norm~\cite{Kim:2024eoa}. Null charges associated with infinite-distance boundaries do not constitute an additional sector of irreducible generators. They may instead arise as total charges of reducible LST sectors or as null accumulation directions of an infinitely generated BPS cone. The former are already contained in the cone generated by their LST components, while the latter are automatically included by our closed-cone convention. Thus the fixed generators together with the remaining negative-norm generators determine the complete closed BPS cone. This closed-generator structure is a physical input of the reconstruction, rather than a consequence of the definition of closure. For $T=1$, where a primitive null charge can itself generate an extremal boundary, the tensor intersection structure is already that of a Hirzebruch surface $\mathbb F_n$ as discussed in~\cite{Kim:2024eoa}.

\subsection{Physical assumptions and scope}
\label{sub:physical-inputs}

Although the charge lattice and anomaly data strictly constrain the allowed string charges, they do not uniquely determine which charges are BPS-populated. Table~\ref{tab:reconstruction-structure} summarizes the lattice data and structural conditions on the BPS cone reviewed above.
Our general reconstruction uses three additional physical inputs:

\begin{table}[htbp]
\centering
\small
\setlength{\tabcolsep}{4pt}
\renewcommand{\arraystretch}{1.15}
\begin{tabular}{@{}>{\raggedright\arraybackslash}p{0.50\textwidth}>{\raggedright\arraybackslash}p{0.46\textwidth}@{}}
\toprule
Structure/condition & Reference \\
\midrule
Charge lattice and anomaly data
& \eqref{eq:anomaly-factorization}, \eqref{eq:b0-square}; Section~\ref{sub:lattice-anomaly-data} \\
\addlinespace
BPS cone and generators
& \eqref{eq:bps-cone-generators}, \eqref{eq:generator-nonpositive-norm}, \eqref{prop:mutual-nonnegativity} \\
\addlinespace
Infinite-distance boundaries governed by little string theories
or critical strings
& Section~\ref{sec:physical-setup} \\
\addlinespace
Causality of the dual tensor cone
& \eqref{eq:dual-cone-causality} \\
\addlinespace
Gravitational anomaly charge: $b_0\in\BPSCone$
& \eqref{eq:b0-cone-membership} \\
\addlinespace
$H$-string and its LST decomposition ($T>1$)
& \eqref{eq:h-string}, \eqref{eq:h-string-lst-decomposition}, \eqref{eq:fixed-set}; Section~\ref{sub:refined-tensor-structure} \\
\addlinespace
Exhaustion of the generators ($T>1$)
& Section~\ref{sub:fixed_generators} \\
\addlinespace
Blowdown consistency
& Boxed statement in Section~\ref{sub:refined-tensor-structure} \\
\bottomrule
\end{tabular}
\caption{Lattice data and structural conditions used in the reconstruction.}
\label{tab:reconstruction-structure}
\end{table}

\begin{itemize}
    \item \textit{Completeness of the SCFT/LST classification.}
    The SCFT and LST classification is assumed complete at the level of tensionless sectors and their BPS string charge spectra, not only primitive generators. In particular, the classified negative-norm generators leave only external E-strings (which carry no gauge algebra) outside the fixed sector; null accumulation directions are included as explained above.
    \item \textit{BPS Ray Completeness Hypothesis.}
    This conjecture, formulated in Section~\ref{sub:bps-ray-completeness}, guarantees that every locally admissible supersymmetric string-source ray contains a nonzero integral charge realized by a BPS configuration.
    \item \textit{Local SCFT description of finite-distance limits.}
    We assume that finite-distance tensionless limits are governed by local SCFTs. Once a candidate ray is shown to be extremal and to support such a boundary, the complete classification identifies its local string sector and hence its primitive populated charge.
\end{itemize}

We keep these physical inputs separate from the numerical candidate analysis. In particular, neither the numerical candidate conditions nor ordinary charge completeness alone implies a BPS population. Nevertheless, in Sections~\ref{sub:lowest-set} and~\ref{sub:null-boundary-selection} as well as in Appendix~\ref{app:T10-reconstruction}, we prove the required BPS ray population for $T\leq10$ directly within EFT. Thus the reconstruction in this range does not rely on BPS Ray Completeness as an independent conjectural assumption. Section~\ref{sec:geometry} separately proves effectivity on an auxiliary rational surface realizing the fixed tensor data and explains how a physical F-theory realization promotes this result to the BPS population.
The exceptional branch $T=9$, $b_0=0$ is discussed separately in Section~\ref{sub:exceptional-T9}.

\section{Compatible E-string candidate sets}
\label{sec:E-string-set}

In the previous section, we identified a fixed set of BPS generators that is completely determined by low-energy EFT data. These generators consist of  primitive charges belonging to the {\it H}-string LST sectors, all external generators supporting gauge algebras, and the M-strings that cannot be removed by a Weyl reflection. The only remaining ambiguity in the reconstruction of the BPS cone therefore lies in the sector of E-strings with positive intersection against the \textit{H}-string. The purpose of this section is to formulate and analyze this residual problem purely within 6d effective field theory. For this reconstruction, we focus on theories with $T>1$, since the tensor sectors with $T\leq1$ have already been fully characterized~\cite{Kim:2024eoa}.

Our strategy proceeds in three steps. First, we systematically enumerate all charges in the tensor charge lattice that satisfy the numerical E-string conditions and intersect every fixed generator non-negatively. Second, we organize these charges into mutually compatible sets. Finally, the $H$-string degree selects a unique lowest-ordered compatible set. The purely EFT step that promotes this set to the physical BPS generators will be presented in the next section.

\subsection{Candidate E-string charges}
\label{sub:candidate-charges}

Once the set of fixed generators $\Fixed$ has been determined, every remaining BPS generator must, as explained above, be an external E-string. Its charge $e$ must therefore satisfy
\begin{equation}
 e^2=-1\,,
 \qquad b_0\cdot e=1\,,
 \qquad e\cdot f>0\,,
 \qquad e\cdot F_a\geq0\quad\text{for every }F_a\in\Fixed\,.
 \label{eq:candidate-set}
\end{equation}
We denote by $\mathcal C_E$ the set of lattice charges $e\in\Gamma$ satisfying~\eqref{eq:candidate-set}. A candidate charge realized as a physical E-string generator must additionally satisfy the following supersymmetric consistency requirements:
\begin{enumerate}
 \item \emph{A tensionless limit.} There must be an accessible finite-distance point in the tensor moduli space at which $J\cdot e=0$. 

 \item \emph{E-string current algebra.} The gauge algebras intersecting $e$ must admit a simultaneous embedding into the level-one $E_8$ current algebra of the E-string~\cite{Heckman:2015bfa,Kim:2014dza}. In particular, the non-Abelian gauge factors occur at the induced levels $k_i=e\cdot b_i$~\cite{Kim:2016foj,Shimizu:2016lbw,Kim:2019vuc}, and must satisfy
 \[
  \bigoplus_{i:\,k_i>0}(\mathfrak g_i)_{k_i}\subseteq(E_8)_1.
 \]
 \item \emph{Intersection bounds.} The intersections of $e$ with the other primitive BPS generators must obey the applicable bounds derived in~\cite{Kim:2026ldd}, with the appropriate self-intersections, gauge algebras, and LST or non-Higgsable sectors taken into account.
\end{enumerate}
These are necessary conditions for a physical generator, not a proof of a BPS population. We call numerical E-string candidates satisfying these additional conditions \emph{locally admissible supersymmetric E-string charges}. 

The first condition holds for any candidate chosen as a BPS generator. See Section~\ref{sub:compatible_candidate_sets} for the proof. In every ordinary F-theory model, however, the E-string current-algebra and intersection constraints are automatically satisfied by consistency of the compactification. Furthermore, in Section~\ref{sec:geometry}, we prove the effectivity of all numerical candidates defined above in geometric models using Riemann-Roch and nefness of $f$. We retain~\eqref{eq:candidate-set} as the definition of the numerical candidate pool and impose the additional requirements when identifying its physical generators. Given the tensor charge lattice and the fixed generators, the enumeration itself remains a purely lattice-theoretic problem.

In the generic case, a candidate charge can be expanded as
\begin{align}
    \label{eq:candidate-ansatz}
    e_i=p f + q F_0-\sum_jn_jx_j \ , \quad p,q,n_j\in \mathbb{Z}_{\ge0}\, ,
\end{align}
where $f$ is the {\it H}-string charge, identified at the level of the lattice with the fiber class of $\mathbb{F}_n$. The charge $F_0$ is one of the fixed generators satisfying $f\cdot F_0=1$, identified with a section class of $\mathbb{F}_n$, while the $x_j$ denote the blowup classes with $x_j^2=-1, b_0\cdot x_j=1$, $f\cdot x_j=F_0\cdot x_j=0$, and $x_j\cdot x_k=0$ for $j\neq k$.

If the original fixed set contains no generator $F_0$ with $f\cdot F_0=1$, we add an E-string charge obeying $f\cdot F_0=1$ to $\Fixed$. The remaining fixed generators $F_a\in \Fixed$ can then be represented by suitable linear combinations of $f$, $F_0$, and  the blowup classes $x_j$ that reproduce their prescribed intersection structure. The existence of such a representation follows from the requirement that the tensor base admits a sequence of blowdowns to the minimal charge lattice of a Hirzebruch surface, as proven in \cite{Kim:2024eoa}, which in this case is $\mathbb F_1$. Otherwise, the fixed set fails the required consistency test and cannot define the tensor sector of a consistent six-dimensional supergravity theory.

The non-negativity of the coefficients in~\eqref{eq:candidate-ansatz} follows from the intersection properties of fixed generators for a given choice of {\it H}-string charge $f$. Each total exceptional charge $x_j$ is a non-negative integral combination of fixed LST generators, so every remaining candidate obeys $e\cdot x_j\geq0$. Writing $F_0^2=-n$ with $n\geq0$, the basis pairings give
\begin{equation*}
 q=e\cdot f>0,\qquad
 n_j=e\cdot x_j\geq0,\qquad
 p=e\cdot F_0+nq\geq0.
\end{equation*}
Here the last inequality uses $F_0\in\Fixed$. Charge integrality then gives $p,q,n_j\in\mathbb Z_{\geq0}$.

We emphasize again that this construction relies only on the EFT result that the intersection structure of the tensor charge lattice admits a realization in terms of the topology of a K\"ahler surface. The identifications of $f$ and $F_0$ with the fiber and section classes of $\mathbb{F}_n$ should therefore be understood only at the level of lattice data. EFT alone does not yet establish that every generator is represented by a holomorphic curve on a K\"ahler surface such as a blowup of $\mathbb{F}_n$.

We may now enumerate all candidate E-string charges satisfying~\eqref{eq:candidate-set}. For this purpose, it is useful to define the \textit{H}-degree
\begin{equation}
    d(e)=f\cdot e\,.
    \label{eq:H-degree}
\end{equation}
This degree is strictly positive for every candidate charge $e$. The search can then be performed systematically degree by degree. One first enumerates all candidates at a fixed small value of $d(e)$ and subsequently increases the degree one unit at a time. For theories with a finite candidate set, including those with $T<9$, this procedure terminates after finitely many steps. If the theory admits infinitely many candidates, as in the ${\rm dP}_9$ case, the search produces an infinite sequence. Nevertheless, the enumeration remains practical order by order because the search at each fixed degree is finite. This property is sufficient for the arguments developed below.

It is important to note that the resulting list contains all physical external E-string generators, but it generally contains additional charges as well. Satisfying the numerical E-string conditions does not ensure that a charge is populated, nor does it guarantee that the charge generates an extremal ray of the BPS cone. The essential task is therefore to extract the genuine BPS generators from this larger candidate set.

\subsection{Compatible candidate sets}
\label{sub:compatible_candidate_sets}

The generators of a BPS cone cannot be selected independently. As stated in \eqref{prop:mutual-nonnegativity}, distinct generators must intersect non-negatively.  The numerical E-string conditions~\eqref{eq:candidate-set}, however, do not ensure mutual non-negativity among candidate charges.  Therefore, if two candidates intersect negatively, they cannot both belong to the set of BPS generators:
\begin{equation}
 e_i\cdot e_j<0
 \quad\Longrightarrow\quad
 e_i\text{ and }e_j\text{ cannot both be BPS generators}.
 \label{eq:negative-edge}
\end{equation}
Consequently, candidate E-string charges must be chosen collectively rather than individually.

We introduce the notion of a compatible candidate set.

\begin{framed}
A compatible candidate set $\mathcal{S}$ is a maximal collection of candidate E-string charges satisfying
\begin{equation}
e_i\cdot e_j\ge0\,,
\qquad
i\neq j\,.
\end{equation}
\end{framed}

Then, let $\mathcal G$ be any set of external E-string generators that, together with the fixed generators, generates a physical BPS cone $\BPSCone$. Since distinct BPS generators have non-negative mutual intersections, $\mathcal G$ must be contained in one of the compatible candidate sets:
\begin{equation}
\mathcal G\subseteq\mathcal S_a
\qquad\text{for some }a.
\end{equation}
We now show that $\mathcal G$ is itself maximal. Suppose, to the contrary, that $\mathcal G\subsetneq\mathcal S_a$ for some maximal compatible set $\mathcal S_a$, and choose
$e'\in\mathcal S_a\setminus\mathcal G$. By the candidate conditions and the compatibility of $\mathcal S_a$,
\begin{equation}
C_i\cdot e'\geq0
\qquad\text{for every generator }C_i\in\Fixed\cup\mathcal G.
\end{equation}
By continuity, the same non-negativity holds on the closed cone generated by these charges, including its accumulation directions. Hence $e'$ belongs to the tensor cone $\BPSCone^\vee$ dual to the physical BPS cone. This tensor cone lies within the closure of the Lorentzian positive cone, so every vector it contains must have non-negative norm. This contradicts the defining E-string condition $e'^2=-1$. Therefore, $\mathcal G$ is maximal and must coincide with one of the compatible sets:
\begin{equation}
\mathcal G=\mathcal S_a\qquad\text{for some }a\,.
\label{eq:unique-set-propagation}
\end{equation}

The external E-string sector is empty if and only if the fixed generators already generate the complete physical BPS cone. To show this, suppose that $\Fixed$ generates the complete closed BPS cone but that an external E-string candidate $e$ nevertheless exists. The candidate conditions give $e\cdot F_a\geq0$ for every $F_a\in\Fixed$. Hence, $e$ is non-negative on the entire physical BPS cone and therefore belongs to its dual tensor cone, $e\in\BPSCone^{\vee}$. But the tensor cone lies within the closure of the Lorentzian forward cone and therefore contains no vector of negative norm, in contradiction with $e^2=-1$. Thus, if the external E-string sector is non-empty, the fixed generators cannot be complete and at least one compatible candidate set must furnish a new extremal BPS generator. 

Conversely, if the external E-string sector is empty, the classification implies that no external negative-norm generator remains outside $\Fixed$. The physical BPS cone is then generated entirely by the fixed set. Therefore,
\begin{equation}
 \{\mathcal S_a\}=\{\varnothing\}
 \quad\Longleftrightarrow\quad
 \mathcal G=\varnothing
 \quad\Longleftrightarrow\quad
 \BPSCone=\operatorname{Cone}(\Fixed)\, .
 \label{eq:emptyset}
\end{equation}

For each compatible set $\mathcal S_a$, define the proposed cone $\mathcal K_a=\operatorname{Cone}(\Fixed\cup\mathcal S_a)$. Every physical BPS cone compatible with the known EFT constraints must coincide with one of these cones:
\begin{equation}
    \BPSCone=\mathcal K_a\qquad\text{for some }a\,.
\label{eq:physical-compatible-cone}
\end{equation}
This does not imply that every $\mathcal K_a$ is physically realized. For a proposed cone $\mathcal K_a$, assume that its tensor moduli space has a nonempty interior. Choose a future-directed timelike interior point $J_0$ of its dual cone, normalized by $J_0^2=1$. Every candidate $e\in\mathcal S_a$ admits a finite-distance wall in the corresponding tensor moduli space. Indeed, the vector $\widetilde J_e=J_0+(J_0\cdot e)e$ obeys $\widetilde J_e\cdot e=0$ and $\widetilde J_e^2=1+(J_0\cdot e)^2>0$, while compatibility gives $\widetilde J_e\cdot C=J_0\cdot C+(J_0\cdot e)(e\cdot C)>0$ for every other generator $C$. Normalizing $\widetilde J_e$ therefore produces the desired wall point, while the normalized line segment joining $J_0$ to $\widetilde J_e$ provides an accessible path that remains inside the dual cone. Consequently, the tensionless-wall condition stated in Section~\ref{sub:candidate-charges} alone neither singles out the physical compatible set nor proves that its candidate charges are populated by BPS strings.

Our reconstruction problem is therefore converted from the selection of individual charges into the selection of one compatible set that defines the physical BPS cone. At this point, however, the pairwise intersection constraints treat all compatible sets on an equal footing and provide no criterion for selecting among them.

\subsection{What compatibility does and does not determine}
\label{sub:compatibility-limit}

The construction above sharply reduces the ambiguity in the E-string sector. Starting from the fixed generators, the numerical E-string conditions produce a candidate spectrum that can be enumerated systematically at each fixed $H$-string degree, while mutual non-negativity organizes these candidates into maximal compatible sets. Moreover, the maximality argument shows that the external E-string generators of any physical BPS cone must coincide with one of these sets. The reconstruction problem has therefore been reduced from determining individual BPS charges to selecting the physical set among a collection of compatible candidate sets.

At this stage, however, the analysis remains purely numerical and does not determine which charge is actually BPS-populated. A lattice charge satisfying all candidate conditions need not be realized by a dynamical BPS string, and mutual compatibility only establishes that a collection of candidates can consistently coexist as primitive generators, but it does not imply that they actually do so. In particular, neither the candidate conditions nor pairwise non-negativity alone distinguish the physical compatible set. This remaining ambiguity will be resolved in the next section.



\section{Uniqueness of the physical BPS cone}
\label{sec:eft}

The remaining reconstruction problem consists of two logically distinct steps: selecting the distinguished set of numerical E-string candidates and establishing that the corresponding charges are physically populated by primitive BPS strings. We first use only the charge lattice with the chosen $H$-string, compatibility, and the positive integral $H$-string degree to define the lowest-ordered compatible set and prove its existence and uniqueness. This part of the argument is purely lattice-theoretic and does not require any assumption about BPS population.

We then turn to the physical question of whether the selected numerical charges are actually carried by BPS strings and can define boundaries of the tensor moduli space. Here we use the fundamental physical input that boundaries of the tensor moduli space are supported by BPS strings that become tensionless there, together with the classification of BPS generators. These boundary strings determine the walls of the physical tensor cone and, consequently, the BPS string cone dual to it. The Lorentzian structure of the tensor moduli space then provides the global constraint needed to determine the fate of the remaining candidate charges. For the general reconstruction, the BPS Ray Completeness Hypothesis, introduced in Section~\ref{sub:bps-ray-completeness}, supplies BPS population along the lowest-ordered charge directions. Together with the classification of the resulting finite-distance tensionless sectors, this determines the complete BPS cone.


\subsection{The \texorpdfstring{$H$}{H}-string ordering of E-string charges}
\label{sub:H-ordering}

We now use the \textit{H}-string to distinguish E-string candidates that cannot simultaneously appear as primitive BPS generators. As introduced in the previous section, every external E-string candidate has positive intersection with the \textit{H}-string. It is therefore natural to associate to each candidate charge $e$ its \textit{H}-string degree
\begin{equation}
 d(e)=f\cdot e>0\,.
 \label{eq:H-degree-ordering}
\end{equation}
Since both $f$ and $e$ are integral charges, $d(e)$ is a positive integer. This provides a natural ordering of the candidate E-string charges that is intrinsic to the 6d EFT.

An important property of this ordering is that two distinct E-string candidates of the same \textit{H}-string degree cannot intersect negatively. To see this, consider two candidates $e$ and $e'$ satisfying
\begin{equation}
 f\cdot e=f\cdot e'\,.
\end{equation}
Their difference $\delta=e-e'$ obeys
\begin{equation}
 f\cdot\delta=0\,,
 \qquad
 b_0\cdot\delta=0\,.
 \label{eq:equal-degree-orthogonality}
\end{equation}
Since $f$ is a null charge in the Lorentzian tensor charge lattice, a non-zero charge orthogonal to $f$ has non-positive norm, with the null direction proportional to $f$. The latter possibility is excluded here because $b_0\cdot f=2$ while $b_0\cdot\delta=0$. Hence, for two distinct candidates,
\begin{equation}
 \delta^2<0.
\end{equation}
Using the E-string conditions $e^2=(e')^2=-1$, this becomes
\begin{equation}
 (e-e')^2=-2-2e\cdot e'<0\,.
\end{equation}
The intersection pairing is integral, and therefore
\begin{equation}
 d(e)=d(e') \quad\Longrightarrow\quad e\cdot e'\geq0\,.
 \label{eq:equal-degree-nonnegative}
\end{equation}
Thus, whenever two E-string candidates intersect negatively, their \textit{H}-string degrees must be different. The \textit{H}-string consequently provides an unambiguous ordering for every negatively intersecting pair.

We next explain the physical significance of this ordering. Consider two E-string candidates satisfying
\begin{equation}
 e\cdot e'=-m<0,
 \qquad
 f\cdot e<f\cdot e',
 \qquad
 m\in\mathbb Z_{>0},
 \label{eq:negative-ordered-pair}
\end{equation}
and suppose that the lower-degree charge $e$ belongs to the physical BPS cone, $e\in\BPSCone$. We claim that the higher-degree charge $e'$ cannot then be a primitive BPS generator. This requires only cone membership, not BPS population of the exact charge $e$.

Suppose, to the contrary, that $e'$ is a BPS generator. Let us define a charge
\begin{equation}
 D=f+d(e')e',
 \qquad D\cdot e'=0.
 \label{eq:ordered-decomposition}
\end{equation}
For every other BPS generator $C$, mutual non-negativity and $f\cdot C\geq0$ give
\begin{equation}
 D\cdot C=f\cdot C+d(e')\,e'\cdot C\geq0.
\end{equation}
The same inequality holds on all accumulation directions by continuity. Hence $D$ is non-negative on the entire BPS cone and therefore
\begin{equation}
 D\in\BPSCone^{\vee}=\mathcal T.
 \label{eq:ordered-coefficient}
\end{equation}
On the other hand,
\begin{equation}
 D\cdot e=d(e)-m\,d(e')<0,
 \label{eq:H-ordering-contradiction}
\end{equation}
which contradicts $e\in\BPSCone$. We therefore obtain the ordering rule
\begin{equation}
 e\cdot e'<0,\quad
 f\cdot e<f\cdot e',\quad
 e\in\BPSCone
 \quad\Longrightarrow\quad
 e'\ {\rm is\ not\ a\ BPS\ generator}.
 \label{eq:H-ordering-rule}
\end{equation}

The ordering rule also determines where the higher-degree candidate lies. It turns out that if $e\in\BPSCone$, then $e'$ belongs to the closed BPS cone but cannot be a generator. To see this, consider
\begin{equation}
 \Delta=e'-e,
 \qquad
 \Delta^2=2(m-1)\geq0,
 \qquad
 f\cdot\Delta>0.
 \label{eq:causal-differences}
\end{equation}
Thus $\Delta$ is future causal. The physical dual-cone condition $\mathcal T=\BPSCone^{\vee}\subseteq\overline{\mathcal C^+}$, together with self-duality of the Lorentz cone, implies $\overline{\mathcal C^+}\subseteq\BPSCone$.
Consequently,
\begin{equation}
 e'=e+\Delta\in\BPSCone.
\end{equation}
Since $e$ and $\Delta$ are non-parallel elements of the closed cone, this
decomposition also shows that $e'$ does not span an extremal ray.


The same future-causal difference gives a purely numerical consequence: transitivity of the negative-pair ordering. Consider three E-string candidates satisfying
\begin{equation}
 e_1\cdot e_2<0,\qquad
 e_2\cdot e_3<0,\qquad
 d(e_1)<d(e_2)<d(e_3).
 \label{eq:transitive-three-candidates}
\end{equation}
By the calculation in~\eqref{eq:causal-differences}, both $\Delta_{12}=e_2-e_1$ and $\Delta_{23}=e_3-e_2$ are future causal. Convexity of the future causal cone implies that their sum $\Delta_{13}=\Delta_{12}+\Delta_{23}=e_3-e_1$ is also future causal.
Hence
\begin{equation}
 e_1\cdot e_3
 =-1-\frac12(e_3-e_1)^2
 \leq-1.
 \label{eq:transitive-negative-intersection}
\end{equation}

Therefore $e_1$ and $e_3$ also intersect negatively, and since $d(e_1)<d(e_3)$, the same ordering relates the two endpoints. In this sense the negative-pair ordering is transitive:
\begin{equation}
 e_1\prec e_2,\qquad e_2\prec e_3
 \qquad\Longrightarrow\qquad
 e_1\prec e_3,
 \label{eq:H-order-transitivity}
\end{equation}
where $e_i\prec e_j$ denotes a negative intersection with $d(e_i)<d(e_j)$.

This transitivity will be important below. It means that successive lower-degree choices cannot lead to mutually conflicting selections. If a candidate has a lower-degree negative partner, following such partners toward smaller degree preserves the negative competition with the original charge. Since the \textit{H}-string degree is a positive integer, this process cannot continue indefinitely toward lower degree.

We emphasize that all statements concerning equal-degree intersections and transitivity are purely properties of the numerical candidate charges. The physical ordering rule~\eqref{eq:H-ordering-rule}, by contrast, requires membership of the lower-degree charge in the physical BPS cone. We now use the numerical part of the ordering to identify a distinguished compatible set, and return to the physical population problem afterward.

\subsection{The lowest-ordered compatible set}

\label{sub:lowest-set}

We now apply the \textit{H}-string ordering to the compatible candidate sets constructed in Section~\ref{sub:compatible_candidate_sets}. Recall that the external E-string generators of any physical BPS cone must coincide with one of these maximal compatible sets. The remaining problem is therefore to determine whether the numerical ordering singles out a distinguished choice among them.

We first define the \textit{lowest-ordered set} as follows:
\begin{framed}
A \textit{lowest-ordered set} $\mathcal S_{\rm low}$ is a maximal compatible set of E-string candidates such that, whenever a charge $e\in\mathcal S_{\rm low}$ intersects another candidate $e'\notin\mathcal S_{\rm low}$ negatively, the charge contained in $\mathcal S_{\rm low}$ has the lower \textit{H}-string degree:
\begin{equation}
 e\cdot e'<0
 \qquad\Longrightarrow\qquad
 f\cdot e<f\cdot e'.
 \label{eq:lowest-set-condition}
\end{equation}
\end{framed}
In other words, whenever two E-string candidates compete through a negative intersection, $\mathcal S_{\rm low}$ retains the candidate with the smaller \textit{H}-string degree. We now show that such a set always exists and is unique.

To establish existence, consider the collection of all E-string candidates that have no negatively intersecting partner of lower \textit{H}-string degree. Equivalently,
\begin{equation}
 \mathcal S_{\rm low}
 =
 \left\{
 e\in\mathcal C_E\ \middle|\
 \nexists\,e'\in\mathcal C_E
 \ \text{with}\ 
 e\cdot e'<0,\quad d(e')<d(e)
 \right\}.
 \label{eq:lowest-set-characterization}
\end{equation}
Any two charges in this collection are mutually compatible. Indeed, if two of them intersected negatively, their degrees could not be equal by~\eqref{eq:equal-degree-nonnegative}, and the higher-degree charge would have a lower-degree negative partner, contradicting its inclusion in the collection.

We next show that this compatible collection is maximal. Consider any candidate $e'$ outside it. By definition, $e'$ has a negatively intersecting partner of lower degree. If this lower-degree charge is not yet in $\mathcal S_{\rm low}$, it in turn has another negatively intersecting partner of still lower degree. Since the degree is a positive integer, this descending sequence must terminate at a charge $e\in\mathcal S_{\rm low}$. The transitivity established in~\eqref{eq:H-order-transitivity} then guarantees
\begin{equation}
 e\cdot e'<0.
\end{equation}
Thus every candidate omitted from $\mathcal S_{\rm low}$ is incompatible with at least one candidate contained in it, and no additional candidate can be added. The collection is therefore maximal. By construction, every negative partner outside the set has higher degree than the charge retained inside, so it also satisfies~\eqref{eq:lowest-set-condition}. This proves the existence of $\mathcal S_{\rm low}$. The same argument applies when the full candidate spectrum is infinite, since every descending sequence in the positive integral degree terminates. If there are no external E-string candidates, the corresponding set is simply empty.

The lowest-ordered set is also unique. Suppose that two distinct maximal compatible sets $\mathcal S_1$ and $\mathcal S_2$ both satisfied the lowest-order condition. Choose $e_1\in\mathcal S_1\setminus\mathcal S_2$. Since $\mathcal S_2$ is maximal, there must be some $e_2\in\mathcal S_2$ with
\begin{equation}
 e_1\cdot e_2<0.
\end{equation}
Their \textit{H}-string degrees cannot be equal. If $d(e_1)<d(e_2)$, then $\mathcal S_2$ contains the higher-degree member of this competing pair and violates~\eqref{eq:lowest-set-condition}. If $d(e_2)<d(e_1)$, the same argument shows that $\mathcal S_1$ violates the condition. Hence two distinct lowest-ordered sets cannot exist. We therefore conclude that
\begin{framed}
\begin{equation}
 \mathcal S_{\rm low}\ \text{exists and is unique}. \nonumber
\end{equation}
\end{framed}

This result is entirely numerical. It uses the tensor charge lattice, the fixed generators, the intersections of the E-string candidates, and their \textit{H}-string degrees, but does not assume that any candidate in $\mathcal S_{\rm low}$ is physically populated.

For theories with $T\leq9$, the Lorentzian structure of the tensor charge lattice allows us to go further. In this range we can show directly that the lowest-ordered set is in fact the only compatible candidate set, that every charge in this set is a physical BPS generator, and consequently that the complete BPS cone is uniquely determined.

Consider two distinct E-string candidates $e$ and $e'$. Their difference satisfies
\begin{equation}
 b_0\cdot(e-e')=0,
 \qquad
 (e-e')^2=-2-2e\cdot e'.
 \label{eq:low-T-candidate-difference}
\end{equation}
For $T<9$, the gravitational charge is timelike,
\begin{equation}
 b_0^2=9-T>0,
\end{equation}
so the subspace orthogonal to $b_0$ is negative definite. Since $e-e'$ is non-zero, $(e-e')^2<0$, and charge integrality implies
\begin{equation}
 e\cdot e'\geq0.
\end{equation}
Thus all E-string candidates are mutually compatible.
The same conclusion holds for $T=9$ with $b_0\neq0$. In this case $b_0$ is null and its orthogonal subspace is negative semidefinite. The only possible exception would occur if $e-e'=\lambda b_0$. However,
\begin{equation}
 0=e^2-(e')^2
 =(e-e')\cdot(e+e')
 =2\lambda,
\end{equation}
where we used $b_0\cdot e=b_0\cdot e'=1$. Hence $\lambda=0$, which would imply $e=e'$. Therefore two distinct candidates again have non-negative mutual intersection.

We have thus shown that for $T\leq9$ with $b_0\neq0$ the complete E-string candidate set $\mathcal C_E$ is the unique maximal compatible set and therefore
\begin{equation}
 \mathcal S_{\rm low}=\mathcal C_E\,.
 \label{eq:low-T-lowest-set}
\end{equation}
Section~\ref{sub:compatible_candidate_sets} showed independently that the physical external E-string generator set $\mathcal G$ must coincide with one of the maximal compatible candidate sets. Since there is only one such set in the present case, we immediately obtain
\begin{equation}
 \mathcal G=\mathcal S_{\rm low}=\mathcal C_E,
 \qquad
 T\leq9,\quad b_0\neq0\,.
 \label{eq:low-T-BPS-generators}
\end{equation}
Thus every admissible E-string candidate in this range is a physical primitive BPS generator. Together with the fixed generators, they determine the complete BPS cone uniquely:
\begin{equation}
 \BPSCone
 =
 {\rm Cone}\!\left(\Fixed\cup\mathcal S_{\rm low}\right),
 \qquad
 T\leq9,\quad b_0\neq0.
 \label{eq:low-T-unique-BPS-cone}
\end{equation}
The tensor moduli space is consequently fixed at the same time by positivity of the tensions of these BPS strings. In particular, no additional conjecture about BPS population is required for the reconstruction in this range.

For $T>9$, this argument no longer applies because $b_0^2<0$. Negatively intersecting E-string candidates can occur, and several maximal compatible sets may coexist. The \textit{H}-string ordering nevertheless continues to select the unique numerical set $\mathcal S_{\rm low}$, as established above. What remains is a genuinely physical question: whether the charges selected in this set are populated by dynamical BPS strings. We now turn to this problem.

\subsection{BPS Ray Completeness Hypothesis and physical population}
\label{sub:bps-ray-completeness}


A central point of our construction is to keep numerical candidacy, BPS population, and primitive extremality logically distinct. The $H$-string ordering addresses the numerical selection problem, while physical population is imposed only afterward to promote the selected charges to actual primitive BPS generators.

To supply this remaining physical input, we introduce the \emph{BPS Ray Completeness Hypothesis}, a BPS refinement of the Completeness Hypothesis. The usual Completeness Hypothesis states that every allowed gauge charge is carried by a dynamical state or extended object~\cite{Polchinski:2003bq,Banks:2010zn,Harlow:2018tng}. It does not require these objects to be BPS. Then supersymmetry suggests a stronger statement for charge directions that are themselves locally consistent with supersymmetric string sources. We formulate this as follows:
\begin{mdframed}[
  style=statementstyle,
  frametitle={BPS Ray Completeness Hypothesis}
]
Every oriented rational charge direction that is locally admissible as a supersymmetric charge source in the effective theory contains a non-zero integral charge realized by a dynamical BPS configuration of the same theory.
\end{mdframed}
Equivalently, if $Q$ is a primitive integral charge defining such an admissible direction, then there exists a positive integer $k$ such that
\begin{equation}
 kQ\qquad\text{is BPS-populated}.
 \label{eq:bps-ray-completeness}
\end{equation}
Here, for a 6d supergravity, the relevant locally admissible directions are those associated with supersymmetric gauge-instanton strings \cite{Duff:1997me} or with E-string candidates satisfying the additional admissibility conditions specified in Section~\ref{sub:candidate-charges}.

The conjecture does not require the primitive charge $Q$ itself to be populated. It requires only one BPS-populated charge along the same oriented charge direction.\footnote{ A potentially related proposal is the massless charge sufficiency conjecture~\cite{Morrison:2021wuv}, which, under suitable positivity conditions, states that massless charged states generate the full gauge charge lattice.} Moreover, the corresponding BPS configuration need not be a single elementary state. It may be a finite aligned multi-charge configuration carrying the total charge $kQ$ and preserving the same supersymmetry.

This distinction is important in comparison with the completeness-based assumptions used in earlier applications of string probes to 6d supergravity~\cite{Kim:2019vuc,Tarazi:2021duw}.\footnote{In particular, Section~3 of~\cite{Tarazi:2021duw} assumes that BPS charges generate the full string charge lattice over $\mathbb Z$. This assumption should be distinguished from population of each individual charge in the charge-by-charge formulation below.} Such arguments motivate the population of BPS string charges from completeness and then use the consistency of their worldsheet theories to constrain the bulk EFT. A charge-by-charge version of BPS completeness, in which every admissible primitive integral charge is required to be BPS-populated, is stronger than what is needed here. In particular, known string constructions can contain holes in the BPS spectrum, as discussed in~\cite{Montero:2022vva}. A primitive lattice charge need not itself be BPS-populated even when a higher positive multiple along the same direction is populated. Such examples violate the charge-by-charge BPS completeness assumption but do not violate the BPS Ray Completeness.

To our knowledge, while the stronger charge-by-charge BPS completeness assumption has known counterexamples, no counterexample is currently known in which a locally admissible supersymmetric string direction contains no BPS configuration at any positive integral multiple. We therefore regard the BPS Ray Completeness Hypothesis as a weaker state-population principle that retains the part of BPS completeness relevant for reconstructing the BPS cone.

It is also important that local admissibility in this conjecture is defined independently of the BPS cone that we are trying to reconstruct. A charge is locally admissible when it satisfies the EFT conditions required of a supersymmetric charge source, including charge quantization, anomaly inflow, and the appropriate worldsheet consistency conditions. We do not assume that the charge already belongs to the physical BPS cone. The conjecture can therefore provide independent information about the BPS population of the numerical candidates rather than assuming the desired cone from the outset.

For the reconstruction, it suffices to apply the conjecture to the external E-string candidates in $\mathcal S_{\rm low}$. By the lowest-order condition, none of them has a negatively intersecting candidate of lower \textit{H}-string degree. While these numerical properties alone do not guarantee all local source consistency conditions, once the remaining local supersymmetry criteria listed in Section~\ref{sub:candidate-charges} are verified, every $e\in\mathcal S_{\rm low}$ defines an admissible E-string direction, and the BPS Ray Completeness implies
\begin{equation}
 \forall\,e\in\mathcal S_{\rm low},
 \qquad
 \exists\,k_e\in\mathbb Z_{>0}
 \quad\text{such that}\quad
 k_e e\ {\rm is\ BPS\!-\!populated}.
 \label{eq:lowest-set-population}
\end{equation}
Since the BPS cone is a real cone, this already places every selected charge inside the physical BPS cone,
\begin{equation}
 e\in\BPSCone,
 \qquad e\in\mathcal S_{\rm low}.
 \label{eq:lowest-set-cone-membership}
\end{equation}

At this stage, however, we have established only a BPS configuration somewhere along the corresponding charge direction. We have not yet shown that the primitive charge $e$ itself is BPS-populated or that it is a primitive BPS generator.

We now show that $e$ must define an extremal direction of the physical BPS cone. If $e$ had non-negative intersection with every actual generator $C_A$, then the same would hold on their null accumulation directions by continuity. Together with the candidate conditions on the fixed sector, this would imply
\begin{equation}
 e\in\BPSCone^{\vee}=\mathcal T,
 \label{eq:lowest-candidate-decomposition}
\end{equation}
in contradiction with $e^2=-1$, since the tensor cone contains no spacelike vectors. Therefore some actual BPS generator $C$ must satisfy
\begin{equation}
 e\cdot C<0.
 \label{eq:negative-physical-generator}
\end{equation}
Equivalently, if a negative pairing were found on a null accumulation direction, continuity would give the same negative pairing with sufficiently nearby normalized primitive generators.

This generator cannot belong to the fixed set, because every candidate in $\mathcal S_{\rm low}$ has non-negative intersection with all fixed generators. The classification therefore identifies $C$ as an external E-string generator. If $C=e$, the desired extremal direction has already been found. Otherwise, $C$ is a distinct candidate negatively intersecting $e\in\mathcal S_{\rm low}$, and the lowest-order condition gives
\begin{equation}
 f\cdot e<f\cdot C.
 \label{eq:lowest-vs-physical-generator}
\end{equation}
But $e\in\BPSCone$ by~\eqref{eq:lowest-set-cone-membership}, so the ordering rule~\eqref{eq:H-ordering-rule} excludes $C$ as a BPS generator. This is a contradiction. Thus only $C=e$ is possible, and every $e\in\mathcal S_{\rm low}$ spans an extremal ray of the physical BPS cone.

Having established that $e$ spans an extremal ray of the physical BPS cone, we apply the wall construction of Section~\ref{sub:compatible_candidate_sets} to this cone. It gives an accessible finite-distance point $J_e$ at which only the ray of $e$ has vanishing tension:
\begin{equation}
 J_e^2=1,
 \qquad
 J_e\cdot e=0.
 \label{eq:E-string-wall}
\end{equation}

We now apply the complete classification of tensionless sectors and their BPS string charge spectra assumed in Section~\ref{sec:physical-setup}. The BPS configuration of charge $ke$, with $k>0$ and $e\in\mathcal S_{\rm low}$, becomes tensionless at this finite-distance wall. At this boundary, only charges parallel to $e$ can become tensionless. The primitive charge satisfies $e^2=-1$ and $b_0\cdot e=1$ and supports no gauge algebra. The classified local tensionless sector is therefore the rank-one E-string theory, whose BPS spectrum contains the fundamental E-string of exact charge $e$ as well as its higher-charge sectors. Consequently, the primitive charge $e$ itself is BPS-populated. Within the complete classification of such finite-distance tensionless sectors assumed in Section~\ref{sec:physical-setup}, there is no distinct local SCFT in which only a multiple $ke$ is realized while the fundamental E-string charge $e$ is absent. This argument neither requires $k=1$ nor assumes that the charge-$ke$ configuration consists of $k$ separated strings. It identifies the local theory containing that configuration and uses its BPS spectrum.

Consequently, every charge in the lowest-ordered set is a physical primitive
BPS generator,
\begin{equation}
 \mathcal S_{\rm low}\subseteq\mathcal G.
 \label{eq:lowest-set-in-generators}
\end{equation}
Section~\ref{sub:compatible_candidate_sets} showed that $\mathcal G$ must
itself be a maximal compatible candidate set. Since
$\mathcal S_{\rm low}$ is already maximal, the inclusion above implies
\begin{equation}
 \mathcal G=\mathcal S_{\rm low}\,.
 \label{eq:physical-lowest-set}
\end{equation}
Thus the BPS Ray Completeness Hypothesis, together with local source admissibility and
the complete classification of 6d SCFTs and LSTs, promotes the unique numerical set selected by the \textit{H}-string ordering to the physical set of primitive E-string generators.

By maximality and lowest ordering, every remaining candidate $e'\notin\mathcal S_{\rm low}$ has a lower-degree negative partner $e\in\mathcal S_{\rm low}\subset\BPSCone$. The ordering argument of Section~\ref{sub:H-ordering} therefore gives
\begin{equation}
 e'\in\BPSCone,
 \qquad
 e'\text{ is non-extremal}.
 \label{eq:higher-candidate-completion}
\end{equation}
Thus the higher candidates do not define additional boundaries of the tensor moduli space. If such a candidate is also locally admissible, BPS Ray Completeness further guarantees a BPS-populated positive integral multiple along the same charge direction.

We have therefore determined both the primitive E-string generators and the role of the remaining numerical candidates. Under the physical inputs stated above, the complete BPS string charge cone is uniquely determined by
\begin{framed}
\begin{equation}
 \BPSCone
 =
 {\rm Cone}\!\left(\Fixed\cup\mathcal S_{\rm low}\right)\,.
 \label{eq:reconstructed-BPS-cone}
\end{equation}
\end{framed}
Here $\operatorname{Cone}$ includes the closure, as defined in Section~\ref{sec:physical-setup}. In particular, any null accumulation directions are already included in~\eqref{eq:reconstructed-BPS-cone} and require no separate enumeration.

The tensor moduli space is determined at the same time by positivity of the tensions of these primitive BPS strings. The fixed generators determine the LST and gauge-supported boundaries, while $\mathcal S_{\rm low}$ supplies the remaining E-string boundaries. Higher numerical E-string candidates do not introduce additional boundaries, but they lie in the closed BPS cone as non-extremal charge directions, and when BPS Ray Completeness applies to them, a positive integral multiple is realized by a BPS configuration.

For $T\leq9$, as shown in the previous subsection, the reconstruction follows directly from the Lorentzian structure of the tensor charge lattice and does not require BPS Ray Completeness as an independent input. For larger $T$, BPS Ray Completeness supplies the general population input needed to promote the numerical lowest-ordered set to the physical BPS spectrum. In the following subsection, we identify an additional class of situations in which the required physical information can be obtained directly from EFT considerations without invoking BPS Ray Completeness as an independent input.

\subsection{Selection from null-boundary constraints}
\label{sub:null-boundary-selection}

The BPS Ray Completeness Hypothesis provides the general population input for the reconstruction discussed above. There are, however, cases in which a lowest-ordered candidate can be identified directly from the structure of the tensor moduli space. We now develop a complementary test based on rational null boundaries. This test relies on the Lorentzian consistency of the physical tensor cone and on the exhaustion of the possible primitive BPS generators by the classified sectors, but does not invoke the BPS Ray Completeness Hypothesis.

The physical input is that a primitive future-null lattice charge $R_0$ contained in the tensor cone $\mathcal T$ defines an infinite-distance boundary governed by a parallel critical string or LST charge\footnote{For a unit timelike interior vector $J_0$, convexity places the normalized path
\begin{equation}
 J(s)=\frac{J_0+sR_0}{\sqrt{1+2s\,J_0\cdot R_0}}\in\mathcal T,
 \qquad J(s)^2=1,\qquad s\geq0,
 \label{eq:null-boundary-path}
\end{equation}
inside the tensor cone. Along this path, $J(s)\cdot R_0=(J_0\cdot R_0)/\sqrt{1+2s\,J_0\cdot R_0}\to0$, while the hyperbolic distance diverges as $s\to\infty$.}
\begin{equation}
 q=nR_0,\qquad n\in\mathbb Z_{>0},\qquad b_0\cdot q\in\{0,2\}.
 \label{eq:rational-null-boundary-input}
\end{equation}
The charge $q$ may be the total little-string charge of a reducible LST sector and need not itself be a primitive BPS generator. This is the rational-null-boundary input used here from the classification reviewed in~\cite{Kim:2024eoa,Hamada:2026zta}.

Consider a lowest-ordered E-string candidate $e$ with $d=f\cdot e$. Associated with it is the future-null charge
\begin{equation}
 R_e=2de+f,\qquad R_e^2=0,\qquad
 b_0\cdot R_e=2d+2,\qquad R_e\cdot e=-d.
 \label{eq:candidate-null-test}
\end{equation}
Its future orientation follows from $f\cdot R_e=2d^2>0$. This charge cannot belong to the physical tensor cone. Indeed, writing $R_e=kR_0$ with $R_0$ primitive, charge integrality implies $k\mid d$ and $k\mid(2d+2)$, hence $k\mid\gcd(d,2)$. Consequently $b_0\cdot R_0=(2d+2)/k>2$, so no positive multiple of $R_0$ can satisfy the null-boundary condition~\eqref{eq:rational-null-boundary-input}. Therefore, independently of whether $e$ is populated,
\begin{equation}
 R_e\notin\mathcal T\,.
 \label{eq:null-test-excluded}
\end{equation}

Since the tensor cone is defined by non-negative intersections with all physical BPS generators, some generator must intersect $R_e$ negatively. If $e$ is a physical generator, it already does so. If it is not, there must be another generator $C$ with $R_e\cdot C<0$. It cannot belong to the fixed set~\eqref{eq:fixed-set}, since $R_e\cdot F_a=2d(e\cdot F_a)+f\cdot F_a\geq0$. Nor can it be future null, since two future-null vectors have non-negative pairing. Our exhaustive enumeration of generators therefore identifies $C$ as another external E-string candidate. Writing $e\cdot C=-m$ and $d_C=f\cdot C$, the lowest-order condition and $R_e\cdot C<0$ require
\begin{equation}
 m\in\mathbb Z_{>0},\qquad d<d_C<2dm.
 \label{eq:null-obstruction-window}
\end{equation}
We call such a candidate an \emph{obstruction witness} for $e$.

The Lorentzian lattice further restricts these witnesses. Since $d_C e-dC$ is orthogonal to $f$, it obeys $(d_C e-dC)^2=-d_C^2-d^2+2dm d_C\leq0$. Equality would require $d_C/d=m+\sqrt{m^2-1}$, which is irrational for $m\geq2$ and equals one for $m=1$, contradicting $d_C>d$. Hence
\begin{equation}
 0<d_C(2dm-d_C)<d^2,\qquad d<d_C<d^2.
 \label{eq:finite-null-obstruction}
\end{equation}
The second bound follows because $2dm-d_C$ is a positive integer. Since $d<d_C<d^2$, only finitely many $H$-string degrees $d_C$ can occur, and the candidate enumeration is finite at every fixed degree. The set of possible obstruction witnesses is therefore finite. In particular, no witness exists for $d=1$, and therefore
\begin{equation}
 d(e)=1\quad\Longrightarrow\quad e\in\mathcal G\,.
 \label{eq:degree-one-forcing}
\end{equation}
Every lowest-ordered E-string candidate of degree one is thus forced to be a physical BPS generator without invoking the BPS Ray Completeness.

The null-boundary constraint therefore provides a finite test of the lowest-ordered candidates, independent of BPS Ray Completeness. It fixes every degree-one candidate and can determine additional generators whenever all of their possible obstruction witnesses are independently excluded. Although this does not establish population of every charge in $\mathcal S_{\rm low}$ at arbitrary degree for general $T$, it provides an independent EFT selection criterion that will be useful in the examples below. For $T=10$, all obstruction witnesses can be excluded at every degree. In Appendix~\ref{app:T10-reconstruction}, we prove the candidate intersection bound $e\cdot e'\geq-1$ and use it to establish $\mathcal G=\mathcal S_{\rm low}$ without invoking BPS Ray Completeness.

\subsection{Exceptional branch \texorpdfstring{$T=9$, $b_0=0$}{T=9, b0=0}}
\label{sub:exceptional-T9}

Theories with $T=9$ and $b_0=0$ have a different physical organization from those discussed above. There is no $H$-string charge satisfying $b_0\cdot f=2$, and hence no distinguished $H$-degree for the reconstruction. The condition $b_0=0$ also sharply restricts the allowed BPS generators. The classification in Table~\ref{tab:bps-generator-classification} leaves only $(-2)$-generators with $b_0\cdot C=0$ and null Type II generators~\cite{Kim:2024eoa,Hamada:2026zta}. The charge-lattice consistency conditions fix the tensor lattice as well. Since $b_0$ is characteristic \cite{Monnier:2018nfs}, its vanishing makes the lattice even, and unimodularity together with signature $(1,9)$ then identify it with $\Gamma=U\oplus E_8(-1)$.

The physical role of a $(-2)$-generator depends on whether it supports a gauge algebra. As discussed above, an ungauged M-string defines a wall separating tensor chambers related by a Weyl reflection. Extending the tensor moduli space across these walls removes the M-string charges from the extremal generator set of the extended BPS cone. By contrast, a $(-2)$-generator supporting a gauge algebra corresponds to an instanton string, and its tensionless wall cannot be removed in this way. We may therefore restrict our attention to $(-2)$-generators that support gauge algebras.

The tensionless sectors associated with the remaining gauge-supporting $(-2)$-generators must themselves admit consistent local descriptions. A connected collection of $(-2)$-generators that becomes simultaneously tensionless must organize into a finite ADE configuration when the local limit is an SCFT, or an affine ADE configuration when it defines an LST. In the latter case, a positive null linear combination of the component charges defines the total little-string charge. These ADE configurations determine the corresponding local tensionless sectors and the associated chamber inequalities.
However, the global tensor cone does not need to coincide with a single finite or affine Weyl chamber. It may also contain infinite-distance boundaries associated with allowed null directions of the charge lattice. Its full structure is therefore determined by the physical walls generated by gauge-supporting $(-2)$-charges, together with the allowed null boundary directions. 

If no gauge-supporting $(-2)$-generators remain after extending through the M-string walls, the tensor cone is the full closed future cone. Its self-duality then fixes the BPS cone as follows:
\begin{equation}
 \mathcal T=\overline{\mathcal C^+}\,,
 \qquad
 \BPSCone=\overline{\mathcal C^+}\,.
 \label{eq:exceptional-T9-causal-cones}
\end{equation}
The remaining extremal directions are null and are associated with Type II strings. The physical tensor moduli space is then given by the unit hyperbolic slice of the future cone, with equivalent descriptions identified under the appropriate charge-lattice dualities.

Ordinary geometric F-theory realizations select precisely this gauge-free branch. In this case, the base is an Enriques surface, whose canonical bundle is nontrivial two-torsion and hence numerically trivial. Since $2K_B\sim0$, the Weierstrass line bundles $\mathcal O_B(-4K_B)$ and $\mathcal O_B(-6K_B)$ are trivial, so their sections are constant on the compact connected base. For a generic smooth elliptic fiber, the discriminant is therefore a nonzero constant, and there is no discriminant divisor supporting a seven-brane gauge algebra.
Consequently, any rational $(-2)$ curves on the base correspond to M-string walls rather than gauge-supporting instanton strings. After extending the tensor moduli space across these walls, the tensor sector is described by the cone in~\eqref{eq:exceptional-T9-causal-cones}. Genus-one fiber directions then realize geometrically the Type II string limits associated with the null boundaries of this cone~\cite{Kim:2024eoa}.

The ordinary Enriques realization has $T=9$, $V=0$, and $H=12$~\cite{Morrison:1996pp,Cheng:2023kxa}, in agreement with gravitational anomaly cancellation. At the level of the continuous gauge sector and the extended tensor cone, it provides the unique ordinary geometric realization of a 6d EFT with $T=9$ and no gauge algebra. The EFT analysis above, however, also allows configurations containing gauge-supporting $(-2)$ sectors, unless these are ruled out by additional physical consistency conditions. Thus the uniqueness of the extended cone without gauge algebra and its Enriques realization in geometry should be distinguished from the broader problem of classifying all consistent EFTs and possible UV completions in this exceptional branch.

\section{EFT reconstruction and examples}
\label{sec:reconstruction}

The results of the preceding sections reduce the reconstruction of the BPS string cone to an explicit procedure based entirely on six-dimensional EFT data. Once an \textit{H}-string is chosen, the BPS generators naturally separate into a fixed sector, determined by the associated LST and gauge data, and a residual sector of external E-string candidates. In Section~\ref{sec:E-string-set}, we developed a systematic procedure for enumerating these candidates and organizing them into maximal compatible sets. The \textit{H}-string ordering introduced in Section~\ref{sec:eft} then selects the unique lowest-ordered set $\mathcal S_{\rm low}$, while the BPS Ray Completeness Hypothesis, once local supersymmetric source admissibility has been established and together with the other physical inputs discussed there, promotes this numerical set to the physical E-string generator set. The complete BPS cone can therefore be reconstructed from intrinsic EFT data without assuming an underlying geometric realization. We now summarize this reconstruction procedure and then illustrate it with several concrete examples.

\subsection{The reconstruction procedure}
\label{sub:reconstruction-procedure}

The required EFT inputs are  the gravitational anomaly charge $b_0$, the gauge anomaly vectors $b_i$ and charged matter content, and the classified SCFT/LST sectors compatible with these data. The reconstruction then proceeds as follows.

\begin{enumerate}

\item Choose an \textit{H}-string charge $f$ satisfying
\begin{equation}
 f^2=0, \qquad b_0\cdot f=2\,.
 \label{eq:reconstruction-H}
\end{equation}
For every theory with  $T\geq1$ and $b_0\neq0$, the existence of such a charge follows from the universal tensor-intersection structure reviewed in Section~\ref{sec:physical-setup}.

\item
Determine all BPS generators orthogonal to $f$. By the containment relation in \cite{Kim:2024eoa}, every such generator $C$ belongs to an LST sector with common little-string charge $f$. The classification of LSTs therefore determines the set $\Fixed_{\mathrm{LST}}(f)$ defined in~\eqref{eq:fixed-lst}, including all mutual intersections among its generators.

\item
Add all external BPS generators with $f\cdot C>0$ that support non-Abelian gauge algebras.  Their charges and intersections are fixed by the anomaly coefficients, the charged matter spectrum, and the intersection bounds in \cite{Kim:2026ldd}.  We denote this set by
\begin{equation}
 \Fixed_{\rm gauge}^{+}(f).
\end{equation}

\item
Treat external M-string charges.  An M-string whose $SU(2)$ flavor symmetry is not gauged is removed by a Weyl reflection. M-strings that are tied to an NHC or whose flavor symmetry is gauged remain as fixed generators. After this step one obtains the complete fixed set\footnote{In the notation of Section~\ref{sub:fixed_generators}, $\Fixed^{+}(f)=\Fixed_{\mathrm{gauge}}^{+}(f)\cup\Fixed_{\mathrm M}(f)$.}
\begin{equation}
 \Fixed
 =
 \Fixed_{\rm LST}
 \cup
 \Fixed_{\rm gauge}^{+}
 \cup
 \Fixed_{\rm M}.
 \label{eq:fixed-total-reconstruction}
\end{equation}
After determining the fixed set, the only undetermined generators lie in the external E-string sector $\mathcal G$.

\item Enumerate the remaining external E-string candidates in increasing $H$-string degree:
\begin{equation}
 e^2=-1,
 \qquad
 b_0\cdot e=1,
 \qquad
 f\cdot e>0,
 \qquad
 e\cdot F_a\geq0
 \quad(F_a\in\Fixed).
 \label{eq:reconstruction-candidates}
\end{equation}
The enumeration is finite at every fixed degree even when the complete candidate spectrum is infinite.

\item
For a conceptual description, consider the maximal subsets of candidate E-string charges whose distinct elements have non-negative mutual intersections:
\begin{equation}
e_i\cdot e_j\geq0\, ,\qquad \text{ for every pair \ } e_i,e_j\in\mathcal S \text{\ with\ } i\neq j.
\end{equation}
Maximality means that no further candidate can be added without violating this condition.  Denote the collection of maximal compatible candidate sets by
\begin{equation}
    \{\mathcal S_a\}\,.
\label{eq:max-compatible-list}
\end{equation}

\item
Use the result of Section~\ref{sub:compatible_candidate_sets}, which states that the external E-string generator set $\mathcal G$ for a physical BPS cone must coincide with one of these compatible sets:
\begin{equation}
\mathcal G=\mathcal S_a
\label{eq:physical-one-of-compatible}
\end{equation}
for some $a$. The reconstruction problem is thereby reduced to selecting one maximal compatible set. This is a characterization of the possible physical sets, not a requirement to enumerate the entire collection.

\item Compare the compatible sets using the \textit{H}-degree ordering. A set $\mathcal S$ passes the lowest-order test if
\begin{equation}
    e\in\mathcal S, \quad e'\notin\mathcal S, \quad e\cdot e'<0 \qquad\Longrightarrow\qquad d(e)<d(e').
\label{eq:reconstruction-lowest-test}
\end{equation}
As proved in Section~\ref{sec:eft}, there is a unique set satisfying this condition:
\begin{equation}
\mathcal S=\mathcal S_{\rm low}.
\end{equation}
In practice, one obtains this set directly by discarding every candidate
with a negatively intersecting partner of lower $H$-string degree.
Enumerating all maximal compatible sets is unnecessary.

\item Establish physical selection. Once local supersymmetric source admissibility
of the selected charges is established, BPS Ray Completeness and the
classification of finite-distance tensionless sectors promote
$\mathcal S_{\rm low}$ to the physical generator set.
For $T\leq9$ with $b_0\neq0$, the unique-maximal-set argument already
suffices under the stated structural inputs. The null-boundary criteria
of Section~\ref{sub:null-boundary-selection} provide another independent
route whenever they exclude every obstruction witness.
For $T=10$, this criterion suffices in general, as shown in Appendix~\ref{app:T10-reconstruction}.
These arguments give
\begin{equation}
    \mathcal G =\mathcal S_{\rm low}\, .
\end{equation}
Combining this sector with the fixed set $\mathcal{F}$ gives the complete closed BPS cone:
\begin{equation}
    \BPSCone = \operatorname{Cone} \bigl(\Fixed\cup\mathcal S_{\rm low}\bigr)\, .
\end{equation}
The closure in $\operatorname{Cone}$ includes any accumulation directions when the generator set is infinite.

\end{enumerate}

If the candidate spectrum is infinite, the reconstruction is carried out degree by degree, with the numerical lowest set determined at each finite degree. Once physical selection has been established, the physical closed cone $\BPSCone$ is obtained as the closure of the union of the resulting increasing sequence of cones. The final step is to verify the global consistency conditions of the tensor moduli space. These include the requirement that the reconstructed tensor base have a topology admitting a sequence of blowdowns to a Hirzebruch surface, as well as the condition that the gravitational charge belong to the physical BPS cone, i.e., $b_0 \in \BPSCone$, and the absence of spacelike directions in the dual tensor cone. This cone-membership condition does not by itself imply population of $b_0$ or of a positive multiple. Its failure nevertheless rules out a physical completion.

\subsection{Examples}
\label{sec:examples}

We illustrate the reconstruction procedure in three complementary settings. The first is a finite low-$T$ example in which the EFT reconstruction reproduces a known geometric Mori cone. The second demonstrates that the same procedure can be applied to a non-geometric EFT, where an infinite candidate spectrum can nevertheless be reconstructed exactly to all degrees. The third provides a genuinely global application to the Swampland program: we show that the reconstructed BPS-cone data rule out a proposed tensor sector that passes all previously known EFT consistency tests.

\subsubsection{Reconstruction in a geometric example}

As our first example, consider F-theory compactified on an elliptically fibered Calabi–Yau threefold whose base is the blowup of $\mathbb{F}_3$ at six generic points, so $B=\mathrm{Bl}_6\mathbb F_3$.
We will demonstrate how the complete Mori cone of $B$ can be recovered using the EFT reconstruction procedure.

Let $s$ denote the $(-3)$-section of the base. This curve supports an $\mathfrak{su}_3$ gauge algebra without charged matter and intersects the fiber class $f$, which represents the \textit{H}-string charge. The six blowups introduce exceptional classes $x_i$. Their intersection data are
\begin{align}
&s^2=-3\,,\quad f^2=0\,,\quad x_i^2=-1\,,
\quad b_0\cdot s=-1\,,\quad b_0\cdot f=2\,,\quad b_0\cdot x_i=1\,,
\nonumber\\
&s\cdot f=1\,,\qquad
s\cdot x_i=f\cdot x_i=0\,,\qquad
x_i\cdot x_j=0
\quad\text{for }i\neq j\,.
\end{align}

Since each $x_i$ is orthogonal to $f$, it belongs to an LST sector with common little-string charge $f$. For every $i$, this sector consists of two E-string generators with charges $x_i$ and $f-x_i$, whose sum is $f$. Thus,
\begin{equation}
\Fixed_{\rm LST}(f)= \{x_i,f-x_i\mid i=1,\ldots,6\}.
\end{equation}
The only curve supporting a gauge algebra is $s$. Since $s\cdot f=1$, it is an external fixed generator. The complete fixed set is therefore
\begin{equation}
    \Fixed = \Fixed_{\rm LST}(f)\cup\Fixed^{+}(f), \qquad \Fixed^{+}(f)=\{s\}.
\end{equation}

Next, using the candidate ansatz~\eqref{eq:candidate-ansatz}, a general numerical external E-string charge can be written as
\begin{align}
    e = af + bs -\sum_i n_i x_i \ .
\end{align}
Then an exhaustive enumeration of lattice vectors satisfying the numerical E-string conditions \eqref{eq:candidate-set} leaves only 16 candidate charges:
\begin{align}
    e_{ijkl} = s+3f -x_i-x_j-x_k-x_l \ {\rm with} \ i<j<k<l \ , \quad e_* = s+4f - \sum_{i}x_i\ .
\end{align}
One can check that all these candidate charges intersect one another non-negatively. They therefore form the unique maximal compatible set $\mathcal S =\{e_{ijkl},e_*\}$. This agrees with the general uniqueness result for $T\leq9$ established in Section~\ref{sub:lowest-set}.

Hence $\mathcal S$ is the physical external E-string generator set, and the reconstructed BPS cone becomes
\begin{equation}
    \BPSCone = \operatorname{Cone}\bigl(\Fixed\cup\mathcal S\bigr)=
  \operatorname{Cone}\{x_i,f-x_i,s,e_{ijkl},e_*\}\, .
\end{equation}
This cone agrees precisely with the Mori cone of $B=\mathrm{Bl}_6\mathbb F_3$ computed in Appendix A of \cite{Jefferson:2018irk}. This example demonstrates that the geometric cone is recovered exactly by the EFT reconstruction.

\subsubsection{The \texorpdfstring{$\mathrm{dP}_{10}$}{dP10} example with a fixed gravitational charge}
\label{sub:dp10-fixed-b0}

We next consider an example that does not admit an ordinary F-theory realization. Despite having an infinite spectrum of E-string candidates, its BPS cone can be reconstructed exactly to all degrees. This example also highlights the distinction between a K\"ahler realization of the tensor data and an ordinary F-theory realization of the gauge sector.

Consider a $T=10$ tensor lattice realized by the ten-point blowup of $\mathbb P^2$, which we denote by $\mathrm{dP}_{10}$ without implying that its anticanonical class is ample. In the standard basis,
\begin{equation}
 H^2=1\,,\quad
 E_i\cdot E_j=-\delta_{ij}\,,\quad
 H\cdot E_i=0\,,
 \quad
 b_0=3H-\sum_{i=1}^{10}E_i\,,
 \quad
 f=H-E_1\,,
 \label{eq:dp10-lattice}
\end{equation}
with $b_0^2=-1$, $f^2=0$, and $b_0\cdot f=2$. Here, we choose $f$ as our {\it H}-string charge.

We now place an $\mathfrak{su}_8$ gauge algebra with one hypermultiplet in the symmetric representation on the charge $b_0$. The gravitational charge $b_0$ then becomes a primitive BPS generator, as summarized in Table~\ref{tab:bps-generator-classification}. As discussed in~\cite{Bhardwaj:2019hhd}, this gauge configuration requires a frozen singularity for its F-theory realization and therefore cannot arise in an ordinary unfrozen F-theory compactification.

This configuration is in fact related to the model in Eq.~(4.30) of~\cite{Hamada:2026zta} by a complex structure deformation of the tensor base, accompanied by Weyl reflections removing the M-strings.  The deformation replaces its ungauged $(-2)$ sector by nine $(-1)-(-1)$ LST sectors, while preserving the $\mathfrak{su}_8$ gauge algebra and its symmetric hypermultiplet.

For the chosen \textit{H}-string charge $f$, the corresponding LST sectors are the nine pairs $\{E_i,f-E_i\}$, with $i=2,\ldots,10$. Each pair consists of two E-strings intersecting once, whose charges sum to the common little-string charge $f$. The fixed set is therefore
\begin{equation}
 \Fixed = \{b_0\} \cup \{E_i,f-E_i\mid i=2,\ldots,10\}.
 \label{eq:dp10-fixed}
\end{equation}

Next, we enumerate the external E-string candidates. An external E-string candidate of {\it H}-string degree $d$ has the form $e=aH-(a-d)E_1-\sum_{i=2}^{10}c_iE_i$. Then the candidate conditions reduce to
\begin{equation}
 0\leq c_i\leq d\,,\quad \sum_{i=2}^{10}c_i=2a+d-1\,,\quad \sum_{i=2}^{10}c_i(d-c_i)=2d^2-d-1\, .
 \label{eq:dp10-candidates}
\end{equation}
These equations allow the candidate spectrum to be enumerated systematically at each degree. For example, the complete set of degree-one candidates is
\begin{equation}
 e_{a,S}=aH-(a-1)E_1-\sum_{i\in S}E_i\,,\quad S\subseteq\{2,\ldots,10\}\,,\quad |S|=2a\,,\quad a=0,\ldots,4\,,
 \label{eq:dp10-degree-one}
\end{equation}
which gives in total 256 candidate charges.

The all-degree structure becomes particularly transparent after shifting each candidate by the gravitational charge. Define
\begin{equation}
v_e=e+b_0\,.
\end{equation}
The shifted charges satisfy
\begin{equation}
v_e^2=b_0\cdot v_e=0\,,
\qquad
f\cdot v_e=d(e)+2>0\,,
\qquad
e\cdot e'=v_e\cdot v_{e'}-1\,.
\label{eq:dp10-null-shift}
\end{equation}
Thus every $v_e$ is future null. Two future-null vectors have non-negative mutual intersection, with equality only when they are parallel. Since the lattice pairing is integral, two distinct E-string candidates can therefore intersect negatively only when their shifted null charges are parallel. In that case,
\begin{equation}
e\cdot e'=-1\,.
\end{equation}

Write the shifted null charge as $v_e=np$, where $p$ is primitive and integral and $n\in\mathbb Z_{>0}$. We first show that this primitive reduction again defines a numerical E-string candidate. For every fixed LST generator $F$, we have
\begin{equation}
 n\,p\cdot F=(e+b_0)\cdot F=e\cdot F+1\geq1.
\end{equation}
Since $p\cdot F$ is integral, this implies $p\cdot F\geq1$. In particular,
for each LST pair $\{E_i,f-E_i\}$,
\begin{equation}
 f\cdot p
 =p\cdot E_i+p\cdot(f-E_i)\geq2.
\end{equation} 
The equality $f\cdot p=2$ is, however, impossible. It would then require $p\cdot E_i=p\cdot(f-E_i)=1$ for all nine LST pairs.
Writing a general integral charge as
\begin{equation}
 p=AH-BE_1-\sum_{i=2}^{10}r_iE_i,
\end{equation}
the first condition gives $r_i=1$ for all $i=2,\ldots,10$. Hence
\begin{equation}
 p=AH-BE_1-\sum_{i=2}^{10}E_i.
\end{equation}
The assumption $f\cdot p=2$ then gives $A-B=2$, while
$b_0\cdot p=0$ gives $3A-B=9$. These two equations imply $2A=7$,
contradicting the integrality of $p$. Therefore
\begin{equation}
 f\cdot p\geq3.
\end{equation}
It follows that $e_1=p-b_0$ is again an external E-string candidate.
Indeed,
\begin{equation}
 e_1^2=-1,\qquad
 b_0\cdot e_1=1,\qquad
 f\cdot e_1=f\cdot p-2>0,
\end{equation}
while for every fixed LST generator $F$,
\begin{equation}
 e_1\cdot F=p\cdot F-1\geq0.
\end{equation}

Every candidate along the same shifted null direction belongs to the tower
\begin{equation}
 e_n=np-b_0=ne_1+(n-1)b_0,\qquad
 d(e_n)=n\bigl(d(e_1)+2\bigr)-2,\qquad n=1,2,\ldots.
 \label{eq:dp10-towers}
\end{equation}
Candidates in the same tower have mutual intersection $-1$, whereas candidates associated with different null directions are compatible. The $H$-string ordering therefore selects the lowest-degree member of every tower, characterized by the primitivity of its shifted null charge:
\begin{equation}
 \mathcal S_{\rm low}
 =
 \{e\in\mathcal C_E\mid e+b_0
 \text{ is primitive in }\Gamma\}\,.
 \label{eq:dp10-lowest}
\end{equation}
This statement applies to every null direction represented by an E-string candidate.

The null-boundary criterion of Section~\ref{sub:null-boundary-selection} allows us to establish the physical selection of all lowest-ordered candidates in this example without assuming BPS Ray Completeness. Suppose that a lowest candidate $e_1$ of degree $d$ were not a physical BPS generator. Then an obstruction witness would have to exist with $d<d_C<2dm$. Any negatively intersecting candidate belongs to the same tower as $e_1$, and hence has intersection $-1$, so $m=1$. On the other hand, the tower structure~\eqref{eq:dp10-towers} implies that the next possible degree satisfies $d_C\geq2d+2$, which is incompatible with the required bound $d_C<2d$. Thus no obstruction witness exists, and every charge in~\eqref{eq:dp10-lowest} is forced to be a physical BPS generator. The higher members of each tower are instead non-extremal, as is also explicit from their positive integral decomposition in~\eqref{eq:dp10-towers}.

Under the stated tensor-sector consistency inputs, the above argument identifies the lowest-ordered set with the physical E-string generator set without invoking the BPS Ray Completeness Hypothesis. The physical closed cone is therefore reconstructed to all degrees as
\begin{equation}
\BPSCone
=\operatorname{Cone}\!\left(
\{b_0\}\cup\{E_i,f-E_i\}_{i=2}^{10}
\cup\{e\in\mathcal C_E\mid e+b_0\text{ primitive}\}
\right)\,.
\label{eq:dp10-bps-cone}
\end{equation}
This gives an exact all-degree reconstruction rather than a finite-degree truncation. Moreover, the gravitational charge automatically belongs to the BPS cone, since $b_0$ itself is one of its primitive generators.

The fixed tensor data admit a simple K\"ahler realization. We choose ten very general points on a smooth plane cubic, so that the resulting blowup contains no irreducible $(-2)$-curves. The strict transform of the cubic is an irreducible genus-one curve in the class $b_0=-K_B$, with $b_0^2=-1$, and therefore spans an extremal Mori direction. The exceptional curves $E_i$ and the strict transforms of the lines through the first and $i$-th points realize the remaining fixed classes $f-E_i$, while the pencil of lines through the first point makes $f$ nef.

This K\"ahler realization of the tensor data should be distinguished from an ordinary F-theory realization of the 6d theory. The $\mathfrak{su}_8$ gauge sector with one symmetric hypermultiplet does not admit an ordinary unfrozen F-theory realization. The corresponding local gauge sector can instead be engineered using frozen F-theory singularities, or equivalently the associated $O7^+$ ingredients, which allow symmetric matter~\cite{Bhardwaj:2018jgp,Bhardwaj:2019hhd}.
A local frozen construction, however, does not by itself establish a compact embedding with the precise gravitational charge, tensor lattice, and fixed generators assumed above. The related compact embedding constraints are discussed in~\cite{MorrisonSung:2023}. We therefore leave the full UV-completion status of this example open. Its purpose here is to demonstrate that the specified tensor and local gauge data admit an explicit all-degree reconstruction of the BPS cone, even though an ordinary F-theory realization is obstructed.

\subsubsection{A \texorpdfstring{$T=11$}{T=11} supergravity and global consistency obstruction}
\label{sub:T11-global-obstruction}

Our final example illustrates a different application of the reconstruction. We consider the proposed $T=11$ supergravity configuration of~\cite{Hamada:2026zta}. Its fixed sector satisfies the local lattice and intersection constraints, and the corresponding E-string candidates can be consistently enumerated degree by degree. Nevertheless, we will show that this fixed sector admits no physical BPS cone compatible with the global consistency conditions of the tensor moduli space. The proposed theory therefore cannot admit a consistent UV completion and is placed in the Swampland by this additional non-perturbative constraint. Our argument uses the null-boundary criterion developed in Section~\ref{sub:null-boundary-selection} and does not rely on the BPS Ray Completeness Hypothesis.

The tensor charge lattice of this model is isomorphic to the intersection lattice of $\mathrm{Bl}_{10}\mathbb F_0$. We denote the two fiber classes by $h$ and $f$, and the exceptional charges by $x_i$, $i=1,\ldots,6$, and $y_a$, $a=1,\ldots,4$, with
\begin{equation}
 h^2=f^2=0,
 \qquad
 h\cdot f=1,
 \qquad
 x_i^2=y_a^2=-1,
 \qquad
 b_0
 =
 2h+2f-\sum_{i=1}^6x_i-\sum_{a=1}^4y_a.
 \label{eq:T11-lattice-data}
\end{equation}
All other pairings between basis vectors vanish. 

The charge $f$ will be used as the \textit{H}-string charge throughout this example. It is the common little-string charge of two distinct LST sectors. A convenient presentation of their primitive generators, following the geometric realization displayed in Fig.~1 of~\cite{Hamada:2026zta}, is
\begin{align}
 \mathcal L_1={}&\{y_4-y_3,\ y_3-y_2,\ y_2,\
 f-y_1-y_2-y_3-y_4,\ y_1\},
 \nonumber\\
 \mathcal L_2={}&\{x_1,\ f-x_1-x_2,\ x_2-x_3-x_6,\ x_6,\
 x_3-x_4-x_6,\ x_4-x_5,\ x_5\}.
 \label{eq:T11-LST-sectors}
\end{align}
The corresponding sequences of self-intersections are
\begin{equation}
 (-2,-2,-1,-4,-1)\,,
 \qquad
 (-1,-2,-3,-1,-3,-2,-1)\,,
\end{equation}
respectively.  In each sector, the {\it H}-string charge $f$ decomposes into the primitive generators as
\begin{align}
f={}&(y_4-y_3)+2(y_3-y_2)+3y_2
+(f-y_1-y_2-y_3-y_4)+y_1,
\nonumber\\
f={}&x_1+(f-x_1-x_2)+(x_2-x_3-x_6)+2x_6
+(x_3-x_4-x_6)+(x_4-x_5)+x_5.
\end{align}
The $(-4)$-generator in $\mathcal L_1$ supports an $\mathfrak{so}_8$ gauge algebra. In $\mathcal L_2$, the four gauge-carrying generators of self-intersections $(-2,-3,-3,-2)$ support, in order, $\mathfrak{su}_2$, $\mathfrak g_2$, $\mathfrak{so}_7$, and $\mathfrak{su}_2$, respectively.

In addition to the two LST sectors, there is a single external generator supporting an $\mathfrak{su}_2$ gauge algebra. We first focus on the $r=1$ configuration of Eq.~(3.6) in~\cite{Hamada:2026zta} and show that it does not admit a consistent physical completion. Its external charge is
\begin{equation}
 C^{\rm ext}
 =
 4h+2f
 -2(x_1+x_2+x_3)
 -x_4-x_5
 -\sum_{a=1}^{4}y_a .
 \label{eq:T11-external-generator}
\end{equation}
A direct computation gives
\begin{equation}
 (C^{\rm ext})^2=-2,
 \qquad
 b_0\cdot C^{\rm ext}=0,
 \qquad
 f\cdot C^{\rm ext}=4.
 \label{eq:T11-external-numerics}
\end{equation}
Then the fixed sector is determined as
\begin{equation}
 \Fixed
 =
 \mathcal L_1
 \cup
 \mathcal L_2
 \cup
 \{C^{\rm ext}\}\,.
 \label{eq:T11-fixed-set}
\end{equation}
The only remaining possible primitive generators are external E-string charges.

\paragraph*{Numerical reconstruction.}
We now reconstruct the residual E-string sector directly. An integral candidate charge of \textit{H}-string degree \(d\) can be written as
\begin{equation}
 e[d;\beta;\mathbf p\mid\mathbf q]
 =dh+\beta f-\sum_{i=1}^{6}p_i x_i-\sum_{a=1}^{4}q_a y_a,
 \qquad d=f\cdot e>0.
 \label{eq:T11-candidate-notation}
\end{equation}
The E-string conditions, together with non-negative intersection with the external generator \(C^{\rm ext}\), impose constraints on the coefficients
\begin{align}
 &2d\beta-\sum_i p_i^2-\sum_aq_a^2=-1\,, \quad 2d+2\beta-\sum_i p_i-\sum_aq_a=1\, ,
 \nonumber\\
 &e\cdot C^{\rm ext}
 =2d+4\beta-2(p_1+p_2+p_3)-p_4-p_5-\sum_aq_a\geq0\, .
 \label{eq:T11-candidate-equations}
\end{align}
Compatibility with the two LST sectors is most conveniently expressed through the intersections with their generators, listed in chain order:
\begin{align}
 e\cdot\mathcal L_1={}&
 (q_4-q_3,\ q_3-q_2,\ q_2,\ d-q_1-q_2-q_3-q_4,\ q_1),
 \nonumber\\
 e\cdot\mathcal L_2={}&
 (p_1,\ d-p_1-p_2,\ p_2-p_3-p_6,\ p_6,
 p_3-p_4-p_6,\ p_4-p_5,\ p_5).
 \label{eq:T11-candidate-LST-pairings}
\end{align}
Each of these pairings must be non-negative. Since the little-string decompositions express $d=f\cdot e$ as a weighted sum of the corresponding non-negative integers, all of these intersection numbers are bounded at fixed $d$. The enumeration of E-string candidates is therefore finite degree by degree.

Let $\mathcal C_d$ denote the degree-$d$ candidate set. Exact enumeration gives
\begin{equation}
 \begin{array}{c|rrrrr}
 d&1&2&3&4&5\\ \hline
 |\mathcal C_d|&9&13&27&34&58
 \end{array}
 \label{eq:T11-candidate-counts}
\end{equation}
As a finite check of the reconstruction, enumeration through degree four gives $83$ candidates. Only three pairs intersect negatively, each relating a degree-one candidate
\begin{equation}
 e_1=h-x_2,\qquad
 e_2=h-x_1,\qquad
 e_3=h+f-x_2-x_3-x_4
\end{equation}
to a degree-four candidate. The $H$-string ordering selects the degree-one member of each pair. This finite computation illustrates the reconstruction. 

\paragraph*{Two forced generators.}
The all-degree obstruction is controlled by two particular lowest-ordered
candidates,
\begin{equation}
 E_0=h-x_2,
 \qquad
 Z=
 3h+2f-x_1-2x_2-2x_3-x_4-x_5-y_3-y_4,
 \label{eq:T11-second-orthogonal-E}
\end{equation}
together with the primitive null charge
\begin{equation}
 R=b_0+E_0+Z.
 \label{eq:T11-R-decomposition}
\end{equation}
They satisfy
\begin{equation}
 E_0^2=Z^2=-1,
 \qquad
 E_0\cdot Z=0,
 \qquad
 R^2=b_0\cdot R=0,
 \qquad
 R\cdot E_0=R\cdot Z=0,
 \qquad
 f\cdot R=6.
 \label{eq:T11-R-properties}
\end{equation}
Their intersections with the fixed sector are non-negative, and
\begin{align}
 R\cdot\mathcal L_1={}&(0,1,1,0,1),
 \nonumber\\
 R\cdot\mathcal L_2={}&(2,0,0,1,0,0,2),
 \qquad
 R\cdot C^{\rm ext}=0.
 \label{eq:T11-R-fixed-r12}
\end{align}
The same null charge also generates an infinite numerical family
\begin{equation}
 E_n=E_0+nR,
 \qquad n\geq0,
 \qquad
 E_n^2=-1,
 \qquad
 b_0\cdot E_n=1,
 \qquad
 f\cdot E_n=1+6n,
 \label{eq:T11-infinite-decomposition}
\end{equation}
whose members have non-negative intersection with every fixed generator and
satisfy $E_r\cdot E_n=-1$ for $r\neq n$. No population of this infinite
family, or any integral LST decomposition of $R$, will be assumed.

The two charges $E_0$ and $Z$ belong to the lowest-ordered set $\mathcal S_{\rm low}$. The first has degree one and is therefore forced to be a BPS generator by~\eqref{eq:degree-one-forcing}. The charge $Z$ has degree three and has non-negative intersection with every degree-one and degree-two candidate. To apply the null-boundary test in Section \ref{sub:null-boundary-selection}, a possible obstruction witness would have to satisfy
\begin{equation}
 3<d_C<9,
 \qquad
 0<d_C(6m-d_C)<9.
\end{equation}
For $m\geq2$ the second inequality has no solution. Hence $m=1$ and only $d_C=4,5$ remain. Direct enumeration at these two degrees gives
\begin{equation}
 Z\cdot C\geq0
 \qquad
 \text{for every }
 C\in\mathcal C_4\cup\mathcal C_5.
 \label{eq:T11-Z-witness-check}
\end{equation}
Thus $Z$ has no obstruction witness and is also forced to be a physical BPS generator. Importantly, neither conclusion uses the BPS Ray Completeness Hypothesis or assumes population of the remaining candidates in the lowest-ordered set.

In any putative physical completion, $E_0,Z\in\mathcal G$.
For every other external E-string generator $C$, mutual non-negativity gives
\begin{equation}
 R\cdot C
 =
 b_0\cdot C+E_0\cdot C+Z\cdot C
 =
 1+E_0\cdot C+Z\cdot C
 \geq1.
 \label{eq:T11-actual-R-face}
\end{equation}
Together with~\eqref{eq:T11-R-fixed-r12}, this places $R$ in the physical
tensor cone and shows that $E_0$ and $Z$ are the only external generators
in $\mathcal G$ orthogonal to $R$.

\paragraph*{Closed-cone obstruction.}
We now impose the global requirement that the gravitational charge belong to the physical BPS cone, $b_0\in \BPSCone$. Because $R$ is future null, this test must take into account possible accumulation directions of the BPS generators. Choose a timelike interior tensor modulus $J_0$ and normalize each primitive generator by its pairing with $J_0$. Since $J_0$ lies in the interior of the tensor cone, this pairing controls a fixed Euclidean norm from below, and the normalized generators therefore have compact closure. Along any unbounded sequence of E-string generators, $J_0\cdot C\to\infty$; since $C^2=-1$, every accumulation point of the normalized sequence is future null. Moreover, a future-null vector orthogonal to $R$ must be parallel to $R$.

The normalized closed BPS cone is the convex hull of the compact closure of these normalized generators. Since $R$ has non-negative intersection with every physical generator and with all of their accumulation points, the face orthogonal to $R$ is generated by those directions on which this intersection vanishes. Defining
\begin{equation}
\Fixed_0=\{F\in\Fixed\mid R\cdot F=0\},
\end{equation}
we therefore obtain
\begin{equation}
\BPSCone\cap R^\perp
\subseteq
\operatorname{Cone}
\bigl(
\Fixed_0\cup\{E_0,Z,R\}
\bigr).
\label{eq:T11-closed-null-face}
\end{equation}
In particular, the possible accumulation direction $R$ is explicitly included in the null face, so this argument does not assume that the BPS cone is finitely generated.

We now introduce a charge
\begin{equation}
 \lambda
 =
 f+x_1-x_2-x_3-x_4-x_5-y_1-y_2+y_3+y_4.
 \label{eq:T11-separating-functional}
\end{equation}
A direct calculation gives
\begin{equation}
 \lambda\cdot E_0
 =
 \lambda\cdot Z
 =
 \lambda\cdot F
 =
 0
 \qquad
 (F\in\Fixed_0),
\end{equation}
while
\begin{equation}
 \lambda\cdot b_0
 =
 \lambda\cdot R
 =
 -1.
\end{equation}
Since $R\cdot b_0=0$, any positive multiple of $b_0$ contained in the
closed BPS cone would have to lie on the zero face~\eqref{eq:T11-closed-null-face}.
Suppose therefore that
\begin{equation}
 m b_0
 =
 \sum_{F\in\Fixed_0}\alpha_FF
 +\beta_0E_0+\beta_ZZ+\gamma R,
 \qquad
 m>0,
 \qquad
 \alpha_F,\beta_0,\beta_Z,\gamma\geq0.
 \label{eq:T11-closed-b0-assumption}
\end{equation}
Pairing with $\lambda$ gives $\gamma=m$. Subtracting $mR$ and using
$b_0-R=-E_0-Z$, we obtain
\begin{equation}
 -m(E_0+Z)
 =
 \sum_{F\in\Fixed_0}\alpha_FF
 +\beta_0E_0+\beta_ZZ.
 \label{eq:T11-closed-face-contradiction}
\end{equation}
The left-hand side has strictly negative pairing with the interior tensor modulus $J_0$, whereas every charge on the right-hand side has non-negative pairing with $J_0$. This is impossible. 

We therefore conclude that $b_0\notin\BPSCone$. Since $\BPSCone$ is a cone, this also excludes every positive multiple of $b_0$. The proposed fixed sector consequently admits no physical completion consistent with the stated tensor-sector inputs. This obstruction holds to all degrees, fully accounts for the possible null accumulation direction $R$, and does not rely on the BPS Ray Completeness Hypothesis.

The same argument also excludes the $r=2$ configuration. In this case, $E_0$ and $Z$ remain forced BPS generators, while $R\cdot C^{\rm ext}_2=2$ removes the external fixed generator from the $R$-orthogonal face without changing the contradiction. For $r>2$, the fixed sector already violates the intersection bound of~\cite{Kim:2026ldd}. Consequently, none of the $T=11$ configurations in Eq.~(3.6) of~\cite{Hamada:2026zta} admits a consistent physical completion given the stated tensor-sector inputs, and thus all of them fall into the Swampland.

This example therefore demonstrates that reconstruction of the complete BPS cone can provide a powerful additional Swampland criterion for six-dimensional supergravity.

\section{Geometric interpretation and realization}
\label{sec:geometry}

The reconstruction developed in Sections~\ref{sec:E-string-set} and~\ref{sec:eft} was formulated entirely within six-dimensional effective field theory. We now explain how the same structure appears when the tensor data are realized by a smooth projective rational surface, as in ordinary F-theory compactifications. Geometry enters only at this stage, as an interpretation and independent check, and is not used as an input in the EFT reconstruction.

\subsection{Geometric interpretation of the tensor charges}

Suppose that the tensor charge lattice is identified with the divisor lattice of a smooth projective rational surface $B$, with
\begin{equation}
  K_B=-b_0\, .
\end{equation}
A physical F-theory realization is a special case of this setup, but the surface statements below require only the corresponding surface and the indicated effective curves. In particular, BPS population and geometric effectivity should not be identified automatically. For an actual F-theory base, an effective holomorphic curve can be wrapped by a D3-brane and gives a BPS string carrying the corresponding tensor charge. On an auxiliary comparison surface, by contrast, effectivity is only a geometric property and need not imply the existence of a physical string.

The EFT blowdown~\eqref{eq:eft-blowdown} has the same action on the charge lattice as the contraction of an exceptional curve. This explains why the tensor lattice naturally acquires the topology familiar from $\mathbb P^2$, $\mathbb F_n$, and their blowups. The converse operation is more restrictive geometrically. An inverse lattice blowup does not by itself guarantee that all prescribed classes are simultaneously represented by irreducible curves, that the required incidence points exist, or that no additional curve generators appear. The lattice reconstruction and the existence of a simultaneous geometric realization should therefore be distinguished.

\subsection{Nefness of the \texorpdfstring{$H$}{H}-string}

There is nevertheless a simple and useful geometric consequence of realizing the fixed LST sector. Suppose that its irreducible curves are $F_a$ and that the chosen $H$-string charge has the physical decomposition
\begin{equation}
    f=\sum_a n_aF_a\,, \quad n_a>0\,, \quad f\cdot F_a=0\,.
 \label{eq:geometric-H-decomposition}
\end{equation}
Then $f$ is nef. Indeed, an irreducible curve equal to one of the $F_a$ has zero intersection with $f$. Any other irreducible curve $C$ has non-negative intersection with every $F_a$, since distinct irreducible curves on a smooth surface have non-negative local intersection multiplicities. Therefore
\begin{equation}
    f\cdot C = \sum_a n_a F_a\cdot C \geq0\,.
\end{equation}
Thus $f$ has non-negative intersection with every irreducible effective curve and is nef. If $f$ itself is represented by an irreducible fiber, the same conclusion follows directly from $f^2=0$ and positivity of intersections with every distinct irreducible curve.

In physical terms, the nefness of the \textit{H}-string charge $f$ implies the existence of an asymptotic infinite-distance boundary of the tensor moduli space, where the projective tensor modulus goes as $J\rightarrow af$ with $a>0$ and the \textit{H}-string tension tends to zero~\cite{Lee:2018urn}. Thus every such geometric realization admits an infinite-distance limit consistent with the Emergent String Conjecture~\cite{Lee:2019xtm}, with the \textit{H}-string playing the role of the emergent tensionless string.

\subsection{Effectivity of the E-string candidates}

The numerical E-string candidates acquire a particularly simple geometric interpretation. On the rational surface $B$, every candidate charge $e$ satisfying
\begin{equation}
    e^2=-1\,, \quad b_0\cdot e=1\,, \quad f\cdot e>0\,
\end{equation}
is represented by an effective divisor class.

To see this, surface Riemann-Roch gives
\begin{equation}
    \chi\bigl(\mathcal O_B(e)\bigr) = \chi(\mathcal O_B) +\frac12 e\cdot(e-K_B) =1\,,
 \label{eq:riemann-roch-population}
\end{equation}
where we used $\chi(\mathcal O_B)=1$ for a rational surface, $K_B=-b_0$, $e^2=-1$, and $b_0\cdot e=1$. On the other hand,
\begin{equation}
    f\cdot(K_B-e) = -2-f\cdot e<0\,.
\end{equation}
Since $f$ is nef, the class $K_B-e$ cannot be effective. Serre duality therefore gives
\begin{equation}
    h^2\bigl(B,\mathcal O_B(e)\bigr) = h^0\bigl(B,\mathcal O_B(K_B-e)\bigr) =0\,.
\end{equation}
Using
\begin{equation}
 \chi\bigl(\mathcal O_B(e)\bigr) = h^0-h^1+h^2 =1\,,
\end{equation}
we obtain
\begin{equation}
    h^0\bigl(B,\mathcal O_B(e)\bigr)=1+h^1\bigl(B,\mathcal O_B(e)\bigr)\geq1\,.
\end{equation}
Thus every numerical E-string candidate $e$ is effective.

For an actual F-theory base, this effectivity statement immediately implies BPS population. A D3-brane wrapped on an effective holomorphic divisor representing $e$ gives a BPS string configuration carrying the same tensor charge. The divisor need not be irreducible: a reducible effective representative corresponds to an aligned multi-string configuration, which is sufficient for the notion of BPS population used in this work. Thus every numerical E-string candidate direction is BPS-populated in an ordinary F-theory realization. In particular, BPS Ray Completeness holds for these candidate directions, and in fact the primitive candidate charge itself is populated.

It is important, however, that effectivity does not imply that $e$ is an irreducible curve or a primitive Mori generator. This distinction provides the geometric counterpart of the candidate-selection problem encountered in the EFT reconstruction.

\subsection{Geometric interpretation of the \texorpdfstring{$H$}{H}-string ordering}

Section~\ref{sub:H-ordering} already establishes the $H$-string ordering and the cone inclusion and non-extremality of higher-degree negative partners. Here we explain their geometric interpretation. Let $e$ and $e'$ be distinct effective E-string candidates with $e\cdot e'<0$, and suppose that $e$ is irreducible. Then every effective representative of $e'$ contains $e$ as a component, so
\begin{equation}
    e'=k e+D\,,   \quad k\in\mathbb Z_{>0},  \quad D\ \text{effective},
\label{eq:geometric-decomposition}
\end{equation}
where $D$ does not contain $e$. Nefness of the $H$-string then gives
\begin{equation}
    f\cdot e' = k\,f\cdot e+f\cdot D \geq f\cdot e\,.
 \label{eq:geometric-degree-order}
\end{equation}
The inequality is strict by the equal-degree result~\eqref{eq:equal-degree-nonnegative}.

For an actual F-theory base, wrapped D3-branes realize this integral decomposition as an aligned multi-string configuration:
\begin{equation}
    e'=k e+D\,, \quad e,D\in\BPSCone\,, \quad D\neq0\,,
 \label{eq:bps-decomposition}
\end{equation}
Section~\ref{sub:H-ordering} already established from the EFT analysis that $e'$ lies in the physical BPS cone but does not define an extremal generator. The additional geometric information from here is that $e'$ is represented by an effective divisor and that its non-extremal decomposition can be realized with positive integral multiplicities.

An irreducible exceptional Mori generator therefore cannot have an effective negative partner of strictly smaller $H$-string degree:
\begin{equation}
    e\ \text{is a Mori generator}, \quad e\cdot e'<0 \quad\Longrightarrow\quad f\cdot e\leq f\cdot e'.
 \label{eq:geometric-generator-criterion}
\end{equation}
The numerical consequences of this ordering are those proved in Section~\ref{sub:lowest-set}.

\subsection{Identification with the geometric Mori cone}

We now consider a realization of the tensor base of the 6d EFT as a rational surface. In such a realization, every irreducible curve of negative self-intersection is either contained in the fixed sector or corresponds to an external E-string candidate. The existence and uniqueness of the lowest-ordered set $\mathcal S_{\rm low}$ were already established in Section~\ref{sub:lowest-set}. It remains to show that its charges are precisely the external E-string generators of the Mori cone.

For $e\in\mathcal S_{\rm low}$, Riemann-Roch provides an effective representative
\begin{equation}
 e=\sum_i n_i C_i,
 \qquad
 n_i\in\mathbb Z_{>0},
\end{equation}
with distinct irreducible components $C_i$. Since
\begin{equation}
 e^2=\sum_i n_i\,e\cdot C_i=-1,
\end{equation}
at least one component satisfies $e\cdot C_i<0$. Distinct irreducible curves have non-negative mutual intersection, so this also implies $C_i^2<0$. The component $C_i$ cannot belong to the fixed sector, since $e$ has non-negative intersection with every fixed generator. The classification of generators therefore identifies $C_i$ as another external E-string candidate. Since $C_i$ is a component of $e$ and $f$ is nef,
\begin{equation}
 d(C_i)\leq d(e).
\end{equation}
If $C_i$ and $e$ were distinct charge classes, their negative intersection would exclude equal degree by~\eqref{eq:equal-degree-nonnegative}, and hence $d(C_i)<d(e)$. This contradicts the defining property of $e\in\mathcal S_{\rm low}$. Hence, one proves that $C_i=e$ as charge classes and $e$ itself is a Mori generator.

Conversely, let $e$ be an external E-string Mori generator. If another candidate $e'$ intersects it negatively, \eqref{eq:geometric-generator-criterion} gives $d(e)<d(e')$. The characterization \eqref{eq:lowest-set-characterization} therefore implies $e\in\mathcal S_{\rm low}$. Hence
\begin{equation}
 \mathcal S_{\rm low}
 =
 \{\text{external E-string Mori generators}\}\,.
\end{equation}
All higher E-string candidates are effective but non-extremal. Together with the fixed generators, the charges in $\mathcal S_{\rm low}$ exhaust the negative-norm extremal generators of the Mori cone. Possible null boundary directions are included automatically through the closure in our definition of $\operatorname{Cone}$.

It follows that whenever the tensor sector admits a realization by a rational surface, the complete Mori cone is generated precisely by the fixed set together with the lowest-ordered E-string charges:
\begin{equation}
 \overline{NE}(B)
 =
 \operatorname{Cone}
 \left(\Fixed\cup\mathcal S_{\rm low}\right)\,.
 \label{eq:Mori-EFT-identification}
\end{equation}

For an actual F-theory base, effective curve classes are populated by wrapped D3-brane BPS configurations. The geometric Mori cone is therefore identified with the physical BPS string cone,
\begin{equation}
 \overline{NE}(B)
 =
 \BPSCone
 =
 \operatorname{Cone}
 \left(\Fixed\cup\mathcal S_{\rm low}\right)\,.
 \label{eq:Mori-BPS-identification}
\end{equation}
Thus an F-theory realization provides an independent geometric proof that the charges in $\mathcal S_{\rm low}$ are precisely the physical external E-string generators. More generally, in the same F-theory, Riemann-Roch ensures BPS population of all numerical E-string candidate charges, possibly through reducible multi-string configurations. Hence BPS Ray Completeness for these candidate directions follows directly in the geometric realization, without invoking the conjectural input of Section~\ref{sub:bps-ray-completeness}. For an auxiliary rational surface, by contrast, the conclusion is the geometric Mori-cone identification~\eqref{eq:Mori-EFT-identification}, without the additional physical interpretation in terms of wrapped D3-branes.

\section{Conclusions}
\label{sec:conclusion}

In this work we developed a bottom-up procedure for reconstructing the BPS string charge cone of six-dimensional $\mathcal N=(1,0)$ supergravity from low-energy EFT data. The central organizing principle is the $H$-string, which separates the negative-norm primitive BPS spectrum into sectors already determined by the classified LST and gauge data and a residual sector of external E-strings supporting no gauge algebra.

We showed that the external E-string candidates can be enumerated directly in the tensor charge lattice and ordered by their intersection with the $H$-string. This ordering selects a unique lowest-ordered compatible set $\mathcal S_{\rm low}$. Subject to the physical population inputs of Section~\ref{sub:bps-ray-completeness}, including BPS Ray Completeness in the general case, these charges, together with the fixed sector, determine the complete BPS cone,
\begin{equation}
    \BPSCone = \operatorname{Cone} \left(\Fixed\cup\mathcal S_{\rm low}\right).
\end{equation}
For $T\leq10$ and in the classes controlled by the null-boundary criterion, the required physical selection follows directly from EFT without invoking BPS Ray Completeness as an independent input. More generally, the BPS Ray Completeness Hypothesis provides the population principle needed for the reconstruction.

A striking consequence is that the familiar geometric structure of 6d string compactifications can be recovered from the non-perturbative EFT data rather than assumed from the outset. Whenever the tensor sector is realized by a rational surface, the $H$-string is nef and Riemann-Roch makes every numerical E-string candidate effective. The $H$-string ordering then identifies the lowest-ordered charges with the remaining negative-norm extremal Mori generators, giving
\begin{equation}
    \overline{NE}(B) = \operatorname{Cone}
\left(\Fixed\cup\mathcal S_{\rm low}\right).
\end{equation}
For an F-theory base, this is precisely the BPS string cone generated by wrapped D3-branes. Thus the K\"ahler geometry of the tensor sector emerges from the intrinsic BPS structure of the six-dimensional EFT.

The reconstruction also provides information beyond the determination of the tensor moduli space. As illustrated by the $T=11$ supergravity example in Section \ref{sub:T11-global-obstruction}, a candidate theory may satisfy the local lattice and anomaly constraints while failing a global condition of the reconstructed BPS cone. In that example, no positive multiple of the gravitational charge $b_0$ can belong to the physical cone, which excludes the proposed tensor sector without invoking BPS Ray Completeness and thereby places the theory in the Swampland. The complete BPS cone can therefore provide an additional non-perturbative Swampland criterion for six-dimensional supergravity.

Several questions remain open. For large $T$, implementing the reconstruction in general requires an exhaustive degree-by-degree enumeration of E-string candidates. Developing more efficient analytic and computational methods for this problem would make the construction practical for much broader classes of tensor lattices. It would also be important to understand BPS Ray Completeness from more fundamental quantum-gravitational principles, or to determine whether global tensor-cone and null-boundary constraints can replace it entirely.

Finally, our reconstruction determines the BPS generators and the closed BPS cone, but does not determine the complete spectrum inside the cone. BPS Ray Completeness requires population only somewhere along each admissible supersymmetric charge direction, and non-extremal charges may be realized by aligned multi-string configurations rather than by single BPS strings. Determining which integral charges support elementary BPS strings, their degeneracies and bound states, and whether a stronger form of BPS completeness follows from six-dimensional EFT remain interesting problems. These questions may clarify how much of the full non-perturbative spectrum, beyond the extremal data governing the tensor moduli space, is already encoded in the low-energy theory.

\acknowledgments
We would like to thank Seongmin Jeon and Cumrun Vafa for helpful discussions.
Y.H.\ was supported in part by JSPS KAKENHI Grant Nos.\ JP24K07035 and JP24KF0167, and by JST BOOST Program Japan Grant No.\ JPMJBY25E1.
H.K. is supported by the National Research Foundation of Korea (NRF) grant funded by the Korean government (MSIT) (2023R1A2C1006542).

\appendix
\section{Reconstruction at \texorpdfstring{$T=10$}{T=10}}
\label{app:T10-reconstruction}

We prove two results that apply to the general $T=10$ reconstruction, not only to the example of Section~\ref{sub:dp10-fixed-b0}. First, any two distinct numerical E-string candidates obey $e\cdot e'\geq-1$. Second, the null-boundary criterion of Section~\ref{sub:null-boundary-selection} forces every candidate in $\mathcal S_{\rm low}$ to be a physical generator. The first statement is purely lattice-theoretic. The second retains the classified-generator structure and the rational-null-boundary input of the main text, but does not use the BPS Ray Completeness Hypothesis. Throughout this appendix, candidates belong to the numerical pool $\mathcal C_E$ defined by~\eqref{eq:candidate-set}.

\subsection{The pairwise intersection bound}
\label{sub:T10-pairwise-bound}

At $T=10$, the gravitational anomaly vector satisfies $b_0^2=9-T=-1$. For any candidate $e$, define $v_e=e+b_0$. The E-string conditions and $b_0\cdot f=2$ give
\begin{equation}
 v_e^2=0,\qquad b_0\cdot v_e=0,\qquad
 f\cdot v_e=f\cdot e+2>0.
 \label{eq:T10-null-shift}
\end{equation}
Thus $v_e$ is nonzero and future null, with the same time orientation as the chosen future-null charge $f$. In signature $(1,10)$, two future-null vectors have non-negative pairing, with equality precisely when they are positively proportional. Consequently,
\begin{equation}
 e\cdot e'=v_e\cdot v_{e'}-1\geq-1.
 \label{eq:T10-pairwise-bound}
\end{equation}
Since the charge lattice is integral, a negative intersection between distinct candidates must equal $-1$. This occurs if and only if $e+b_0$ and $e'+b_0$ lie on the same future-null ray. No choice of the fixed set, beyond the candidate conditions, enters this bound, and neither candidate is assumed to be BPS-populated.

\subsection{Selection without BPS Ray Completeness}
\label{sub:T10-null-selection}

We now retain the physical generator structure of Section~\ref{sub:fixed_generators} and the rational-null-boundary condition~\eqref{eq:rational-null-boundary-input}. In particular, the fixed set is taken to consist of negative-norm primitive generators. The negative-norm entries of Table~\ref{tab:bps-generator-classification} satisfy
\begin{equation}
 b_0\cdot F_a\leq1\qquad\text{for every }F_a\in\Fixed.
 \label{eq:T10-fixed-anomaly-bound}
\end{equation}
This includes the frozen generator with $C^2=-1, b_0\cdot C=-1$. 

Let $e\in\mathcal S_{\rm low}$ and write $d=f\cdot e$. The argument of Section~\ref{sub:null-boundary-selection} excludes $R_e=2de+f$ from the physical tensor cone independently of whether $e$ is populated. If $e$ were not a physical generator, another physical external E-string generator $C$ would therefore have to satisfy the obstruction condition~\eqref{eq:null-obstruction-window}. By~\eqref{eq:T10-pairwise-bound}, its negative intersection with $e$ must be $-1$, so the obstruction window becomes
\begin{equation}
 e\cdot C=-1,\qquad d<d_C<2d,\qquad d_C=f\cdot C.
 \label{eq:T10-witness-window}
\end{equation}
We show that no numerical candidate $C$ can satisfy these conditions.

Consider the integral charge $\widetilde e=2e-C$. Direct calculation gives
\begin{equation}
 \begin{gathered}
 \widetilde e=2e-C,\qquad
 \widetilde e^{\,2}=-1,\qquad b_0\cdot\widetilde e=1,\\
 0<f\cdot\widetilde e=2d-d_C<d,\qquad
 e\cdot\widetilde e=-1.
 \end{gathered}
 \label{eq:T10-reflected-candidate}
\end{equation}
It remains to check non-negative intersection with the fixed set. Since $e\cdot C=-1$, the shifted null charges $e+b_0$ and $C+b_0$ lie on the same ray. Pairing their proportionality relation with $f$ determines
\begin{equation}
 C+b_0=(1+t)(e+b_0),\qquad
 t=\frac{d_C-d}{d+2},\qquad 0<t<1.
 \label{eq:T10-segment-parameter}
\end{equation}
The strict inequalities follow from $d<d_C<2d$. Hence
\begin{equation}
 \widetilde e=(1-t)e+t(-b_0).
 \label{eq:T10-reflected-segment}
\end{equation}
In other words, $\widetilde e$ lies strictly between $e$ and $-b_0$ on their affine line. This is a statement about charge vectors, not an assumption that $-b_0$ is physical. For every fixed generator, the candidate inequality $e\cdot F_a\geq0$ and~\eqref{eq:T10-fixed-anomaly-bound} imply
\begin{equation}
 \widetilde e\cdot F_a
 =(1-t)(e\cdot F_a)-t(b_0\cdot F_a)
 \geq-t>-1.
 \label{eq:T10-fixed-sieve}
\end{equation}
Because $\widetilde e$ and $F_a$ are lattice charges, their pairing is an integer and must therefore be non-negative. Thus $\widetilde e\in\mathcal C_E$. But~\eqref{eq:T10-reflected-candidate} makes it a lower-degree candidate negatively intersecting $e$, contradicting the characterization~\eqref{eq:lowest-set-characterization} of $\mathcal S_{\rm low}$.

There is consequently no obstruction witness for any $e\in\mathcal S_{\rm low}$ at $T=10$. Every such $e$ must be a physical primitive generator, so $\mathcal S_{\rm low}\subseteq\mathcal G$. Since $\mathcal G$ is a compatible subset of $\mathcal C_E$ and $\mathcal S_{\rm low}$ is already maximal, this inclusion is an equality. The closed-generator structure then yields
\begin{equation}
 \mathcal G=\mathcal S_{\rm low},\qquad
 \BPSCone=\operatorname{Cone}\!\left(\Fixed\cup\mathcal S_{\rm low}\right) \qquad {\rm for}\quad T=10\,.
 \label{eq:T10-reconstructed-cone}
\end{equation}
The cone includes the closure, as in the main text. Unlike the $T\leq9$ result, this does not identify $\mathcal G$ with the entire candidate pool. Non-extremal higher-degree E-string candidates may still exist. Additionally, the argument establishes neither the population of every non-extremal charge ray nor a physical completion for arbitrary given fixed data. It determines the generators of any physical completion satisfying the stated inputs. In particular, if a forced candidate fails the required worldsheet consistency conditions, those fixed data admit no such completion. The BPS Ray Completeness Hypothesis is not needed for this $T=10$ reconstruction, whereas the generator classification and the rational-null-boundary input remain essential.

\bibliographystyle{JHEP}
\bibliography{references}

@article{Morrison:2021wuv,
    author = "Morrison, David R. and Taylor, Washington",
    title = "{Charge completeness and the massless charge lattice in F-theory models of supergravity}",
    eprint = "2108.02309",
    archivePrefix = "arXiv",
    primaryClass = "hep-th",
    reportNumber = "MIT-CTP-5172, UCSB-MATH-2021-03",
    doi = "10.1007/JHEP12(2021)040",
    journal = "JHEP",
    volume = "12",
    pages = "040",
    year = "2021"
}

@article{Lee:2018urn,
    author = "Lee, Seung-Joo and Lerche, Wolfgang and Weigand, Timo",
    title = "{Tensionless Strings and the Weak Gravity Conjecture}",
    eprint = "1808.05958",
    archivePrefix = "arXiv",
    primaryClass = "hep-th",
    reportNumber = "CERN-TH-2018-190",
    doi = "10.1007/JHEP10(2018)164",
    journal = "JHEP",
    volume = "10",
    pages = "164",
    year = "2018"
}

@article{Monnier:2018nfs,
    author = "Monnier, Samuel and Moore, Gregory W.",
    title = "{Remarks on the Green{\textendash}Schwarz Terms of Six-Dimensional Supergravity Theories}",
    eprint = "1808.01334",
    archivePrefix = "arXiv",
    primaryClass = "hep-th",
    doi = "10.1007/s00220-019-03341-7",
    journal = "Commun. Math. Phys.",
    volume = "372",
    number = "3",
    pages = "963--1025",
    year = "2019"
}

@article{Duff:1997me,
    author = "Duff, M. J. and Liu, James T. and Lu, Hong and Pope, C. N.",
    editor = "Kursunoglu, B. N. and Mintz, S. L. and Perlmutter, A.",
    title = "{Gauge dyonic strings and their global limit}",
    eprint = "hep-th/9711089",
    archivePrefix = "arXiv",
    reportNumber = "CTP-TAMU-46-97, RU-97-7-B, LPTENS-97-52",
    doi = "10.1016/S0550-3213(98)00367-8",
    journal = "Nucl. Phys. B",
    volume = "529",
    pages = "137--156",
    year = "1998"
}

@article{Cheng:2023kxa,
    author = "Cheng, Peng and Melnikov, Ilarion V. and Minasian, Ruben",
    title = "{Flat F-theory and friends}",
    eprint = "2306.00865",
    archivePrefix = "arXiv",
    primaryClass = "hep-th",
    doi = "10.1007/JHEP01(2024)027",
    journal = "JHEP",
    volume = "01",
    pages = "027",
    year = "2024"
}

@article{Ganor:1996mu,
    author = "Ganor, Ori J. and Hanany, Amihay",
    title = "{Small E(8) instantons and tensionless noncritical strings}",
    eprint = "hep-th/9602120",
    archivePrefix = "arXiv",
    reportNumber = "IASSNS-HEP-96-12, PUPT-1595",
    doi = "10.1016/0550-3213(96)00243-X",
    journal = "Nucl. Phys. B",
    volume = "474",
    pages = "122--140",
    year = "1996"
}

@article{Seiberg:1996vs,
    author = "Seiberg, N. and Witten, Edward",
    title = "{Comments on string dynamics in six-dimensions}",
    eprint = "hep-th/9603003",
    archivePrefix = "arXiv",
    reportNumber = "RU-96-12, IASSNS-HEP-96-19",
    doi = "10.1016/0550-3213(96)00189-7",
    journal = "Nucl. Phys. B",
    volume = "471",
    pages = "121--134",
    year = "1996"
}

@article{Jefferson:2018irk,
    author = "Jefferson, Patrick and Katz, Sheldon and Kim, Hee-Cheol and Vafa, Cumrun",
    title = "{On Geometric Classification of 5d SCFTs}",
    eprint = "1801.04036",
    archivePrefix = "arXiv",
    primaryClass = "hep-th",
    doi = "10.1007/JHEP04(2018)103",
    journal = "JHEP",
    volume = "04",
    pages = "103",
    year = "2018"
}

@article{Kumar:2010ru,
    author = "Kumar, Vijay and Morrison, David R. and Taylor, Washington",
    title = "{Global aspects of the space of 6D N = 1 supergravities}",
    eprint = "1008.1062",
    archivePrefix = "arXiv",
    primaryClass = "hep-th",
    reportNumber = "MIT-CTP-4132, NSF-KITP-10-096, UCSB-MATH-2010-19",
    doi = "10.1007/JHEP11(2010)118",
    journal = "JHEP",
    volume = "11",
    pages = "118",
    year = "2010"
}

@article{Kim:2026ldd,
    author = "Kim, Hee-Cheol and Xu, Kai",
    title = "{Intersection Bounds for BPS Strings in Six-Dimensional Supergravity}",
    eprint = "2607.26197",
    archivePrefix = "arXiv",
    primaryClass = "hep-th",
    month = "7",
    year = "2026"
}

@article{Lee:2019skh,
    author = "Lee, Seung-Joo and Weigand, Timo",
    title = "{Swampland Bounds on the Abelian Gauge Sector}",
    eprint = "1905.13213",
    archivePrefix = "arXiv",
    primaryClass = "hep-th",
    reportNumber = "CERN-TH-2019-082",
    doi = "10.1103/PhysRevD.100.026015",
    journal = "Phys. Rev. D",
    volume = "100",
    number = "2",
    pages = "026015",
    year = "2019"
}

@article{MorrisonSung:2023,
    author = "Morrison, David R. and Sung, Benjamin",
    title = "{On the frozen F-theory landscape}",
    eprint = "2310.11432",
    archivePrefix = "arXiv",
    primaryClass = "hep-th",
    doi = "10.1007/JHEP05(2024)126",
    journal = "JHEP",
    volume = "05",
    pages = "126",
    year = "2024"
}

@article{Haghighat:2015ega,
    author = "Haghighat, Babak and Murthy, Sameer and Vafa, Cumrun and Vandoren, Stefan",
    title = "{F-Theory, Spinning Black Holes and Multi-string Branches}",
    eprint = "1509.00455",
    archivePrefix = "arXiv",
    primaryClass = "hep-th",
    doi = "10.1007/JHEP01(2016)009",
    journal = "JHEP",
    volume = "01",
    pages = "009",
    year = "2016"
}

@article{Hamada:2023zol,
    author = "Hamada, Yuta and Loges, Gregory J.",
    title = "{Towards a complete classification of 6D supergravities}",
    eprint = "2311.00868",
    archivePrefix = "arXiv",
    primaryClass = "hep-th",
    reportNumber = "KEK-TH-2572",
    doi = "10.1007/JHEP02(2024)095",
    journal = "JHEP",
    volume = "02",
    pages = "095",
    year = "2024"
}

@article{Romans:1986er,
    title = {Self-duality for interacting fields: Covariant field equations for six-dimensional chiral supergravities},
    journal = {Nuclear Physics B},
    volume = {276},
    number = {1},
    pages = {71-92},
    year = {1986},
    issn = {0550-3213},
    doi = {https://doi.org/10.1016/0550-3213(86)90016-7},
    url = {https://www.sciencedirect.com/science/article/pii/0550321386900167},
    author = {L.J. Romans}
}

@article{Erler:1993zy,
    author = "Erler, Jens",
    title = "{Anomaly cancellation in six-dimensions}",
    eprint = "hep-th/9304104",
    archivePrefix = "arXiv",
    journal = "J. Math. Phys.",
    volume = "35",
    pages = "1819--1833",
    year = "1994"
}

@article{Lee:2019xtm,
    author = "Lee, Seung-Joo and Lerche, Wolfgang and Weigand, Timo",
    title = "{Emergent strings from infinite distance limits}",
    eprint = "1910.01135",
    archivePrefix = "arXiv",
    primaryClass = "hep-th",
    doi = "10.1007/JHEP02(2022)190",
    journal = "JHEP",
    volume = "02",
    pages = "190",
    year = "2022"
}

@article{Cheung:2016wjt,
    author = "Cheung, Clifford and Remmen, Grant N.",
    title = "{Positivity of Curvature-Squared Corrections in Gravity}",
    eprint = "1608.02942",
    archivePrefix = "arXiv",
    primaryClass = "hep-th",
    doi = "10.1103/PhysRevLett.118.051601",
    journal = "Phys. Rev. Lett.",
    volume = "118",
    number = "5",
    pages = "051601",
    year = "2017"
}

@article{Hamada:2018dde,
    author = "Hamada, Yuta and Noumi, Toshifumi and Shiu, Gary",
    title = "{Weak Gravity Conjecture from Unitarity and Causality}",
    eprint = "1810.03637",
    archivePrefix = "arXiv",
    primaryClass = "hep-th",
    doi = "10.1103/PhysRevLett.123.051601",
    journal = "Phys. Rev. Lett.",
    volume = "123",
    number = "5",
    pages = "051601",
    year = "2019"
}

@article{Kim:2016foj,
    author = "Kim, Hee-Cheol and Kim, Seok and Park, Jaemo",
    title = "{6d strings from new chiral gauge theories}",
    eprint = "1608.03919",
    archivePrefix = "arXiv",
    primaryClass = "hep-th",
    reportNumber = "SNUTP16-002",
    month = "8",
    year = "2016"
}

@article{Kim:2014dza,
    author = "Kim, Joonho and Kim, Seok and Lee, Kimyeong and Park, Jaemo and Vafa, Cumrun",
    title = "{Elliptic Genus of E-strings}",
    eprint = "1411.2324",
    archivePrefix = "arXiv",
    primaryClass = "hep-th",
    reportNumber = "SNUTP14-010, KIAS-P14063",
    doi = "10.1007/JHEP09(2017)098",
    journal = "JHEP",
    volume = "09",
    pages = "098",
    year = "2017"
}

@article{Vafa:2005ui,
    author = "Vafa, Cumrun",
    title = "{The String Landscape and the Swampland}",
    eprint = "hep-th/0509212",
    archivePrefix = "arXiv",
    year = "2005"
}

@article{Agmon:2022thq,
    author = "Agmon, Nathan Benjamin and Bedroya, Alek and Kang, Monica Jinwoo and Vafa, Cumrun",
    title = "{Lectures on the String Landscape and the Swampland}",
    eprint = "2212.06187",
    archivePrefix = "arXiv",
    primaryClass = "hep-th",
    month = "12",
    year = "2022"
}

@article{Morrison:1996pp,
    author = "Morrison, David R. and Vafa, Cumrun",
    title = "{Compactifications of F-theory on Calabi--Yau threefolds. II}",
    eprint = "hep-th/9603161",
    archivePrefix = "arXiv",
    journal = "Nucl. Phys. B",
    volume = "476",
    pages = "437--469",
    year = "1996"
}

@article{Heckman:2015bfa,
    author = "Heckman, Jonathan J. and Morrison, David R. and Rudelius, Tom and Vafa, Cumrun",
    title = "{Atomic Classification of 6D SCFTs}",
    eprint = "1502.05405",
    archivePrefix = "arXiv",
    primaryClass = "hep-th",
    doi = "10.1002/prop.201500024",
    journal = "Fortsch. Phys.",
    volume = "63",
    pages = "468--530",
    year = "2015"
}

@article{Bhardwaj:2015xxa,
    author = "Bhardwaj, Lakshya",
    title = "{Classification of 6d $ \mathcal{N}=\left(1,0\right) $ gauge theories}",
    eprint = "1502.06594",
    archivePrefix = "arXiv",
    primaryClass = "hep-th",
    doi = "10.1007/JHEP11(2015)002",
    journal = "JHEP",
    volume = "11",
    pages = "002",
    year = "2015"
}

@article{Bhardwaj:2015oru,
    author = "Bhardwaj, Lakshya and Del Zotto, Michele and Heckman, Jonathan J. and Morrison, David R. and Rudelius, Tom and Vafa, Cumrun",
    title = "{F-theory and the Classification of Little Strings}",
    eprint = "1511.05565",
    archivePrefix = "arXiv",
    primaryClass = "hep-th",
    doi = "10.1103/PhysRevD.93.086002",
    journal = "Phys. Rev. D",
    volume = "93",
    number = "8",
    pages = "086002",
    year = "2016",
    note = "[Erratum: Phys.Rev.D 100, 029901 (2019)]"
}

@article{Bhardwaj:2019hhd,
    author = "Bhardwaj, Lakshya",
    title = "{Revisiting the classifications of 6d SCFTs and LSTs}",
    eprint = "1903.10503",
    archivePrefix = "arXiv",
    primaryClass = "hep-th",
    doi = "10.1007/JHEP03(2020)171",
    journal = "JHEP",
    volume = "03",
    pages = "171",
    year = "2020"
}

@article{Bhardwaj:2018jgp,
    author = "Bhardwaj, Lakshya and Morrison, David R. and Tachikawa, Yuji and Tomasiello, Alessandro",
    title = "{The frozen phase of F-theory}",
    eprint = "1805.09070",
    archivePrefix = "arXiv",
    primaryClass = "hep-th",
    reportNumber = "IPMU-18-0001, USCB Math 2018-16, USCB-MATH-2018-16",
    doi = "10.1007/JHEP08(2018)138",
    journal = "JHEP",
    volume = "08",
    pages = "138",
    year = "2018"
}

@article{Kim:2024eoa,
    author = "Kim, Hee-Cheol and Vafa, Cumrun and Xu, Kai",
    title = "{Finite landscape of 6d N=(1,0) supergravity}",
    eprint = "2411.19155",
    archivePrefix = "arXiv",
    primaryClass = "hep-th",
    doi = "10.21468/SciPostPhys.20.1.016",
    journal = "SciPost Phys.",
    volume = "20",
    number = "1",
    pages = "016",
    year = "2026"
}

@article{Hamada:2026zta,
    author = "Hamada, Yuta and Jeon, Seongmin and Kim, Hee-Cheol",
    title = "{6d Supergravity Blocks}",
    eprint = "2607.05496",
    archivePrefix = "arXiv",
    primaryClass = "hep-th",
    reportNumber = "KEK-TH-2849, RIKEN-iTHEMS-Report-26",
    month = "7",
    year = "2026"
}

@article{Morrison:2012np,
    author = "Morrison, David R. and Taylor, Washington",
    title = "{Classifying bases for 6D F-theory models}",
    eprint = "1201.1943",
    archivePrefix = "arXiv",
    primaryClass = "hep-th",
    reportNumber = "UCSB-MATH-2012-01, MIT-CTP-4339",
    doi = "10.2478/s11534-012-0065-4",
    journal = "Central Eur. J. Phys.",
    volume = "10",
    pages = "1072--1088",
    year = "2012"
}

@article{Seiberg:2011dr,
    author = "Seiberg, Nathan and Taylor, Washington",
    title = "{Charge Lattices and Consistency of 6D Supergravity}",
    eprint = "1103.0019",
    archivePrefix = "arXiv",
    primaryClass = "hep-th",
    reportNumber = "MIT-CTP-4216",
    doi = "10.1007/JHEP06(2011)001",
    journal = "JHEP",
    volume = "06",
    pages = "001",
    year = "2011"
}

@article{Monnier:2017oqd,
    author = "Monnier, Samuel and Moore, Gregory W. and Park, Daniel S.",
    title = "{Quantization of anomaly coefficients in 6D $\mathcal{N}=(1,0)$ supergravity}",
    eprint = "1711.04777",
    archivePrefix = "arXiv",
    primaryClass = "hep-th",
    doi = "10.1007/JHEP02(2018)020",
    journal = "JHEP",
    volume = "02",
    pages = "020",
    year = "2018"
}

@article{Green:1984sg,
    author = "Green, Michael B. and Schwarz, John H.",
    title = "{Anomaly Cancellation in Supersymmetric D=10 Gauge Theory and Superstring Theory}",
    reportNumber = "CALT-68-1182",
    doi = "10.1016/0370-2693(84)91565-X",
    journal = "Phys. Lett. B",
    volume = "149",
    pages = "117--122",
    year = "1984"
}

@article{Sagnotti:1992qw,
    author = "Sagnotti, Augusto",
    title = "{A Note on the Green-Schwarz mechanism in open string theories}",
    eprint = "hep-th/9210127",
    archivePrefix = "arXiv",
    reportNumber = "ROM2F-92-49",
    doi = "10.1016/0370-2693(92)90682-T",
    journal = "Phys. Lett. B",
    volume = "294",
    pages = "196--203",
    year = "1992"
}

@article{Kim:2019vuc,
    author = "Kim, Hee-Cheol and Shiu, Gary and Vafa, Cumrun",
    title = "{Branes and the Swampland}",
    eprint = "1905.08261",
    archivePrefix = "arXiv",
    primaryClass = "hep-th",
    journal = "Phys. Rev. D",
    volume = "100",
    number = "6",
    pages = "066006",
    year = "2019"
}

@article{Shimizu:2016lbw,
    author = "Shimizu, Hiroyuki and Tachikawa, Yuji",
    title = "{Anomaly of strings of 6d $\mathcal{N}=(1,0)$ theories}",
    eprint = "1608.05894",
    archivePrefix = "arXiv",
    primaryClass = "hep-th",
    journal = "JHEP",
    volume = "11",
    pages = "165",
    year = "2016"
}

@article{Tarazi:2021duw,
    author = "Tarazi, Houri-Christina and Vafa, Cumrun",
    title = "{On the finiteness of 6d supergravity Landscape}",
    eprint = "2106.10839",
    archivePrefix = "arXiv",
    primaryClass = "hep-th",
    doi = "10.1007/JHEP05(2026)007",
    journal = "JHEP",
    volume = "05",
    pages = "007",
    year = "2026"
}

@article{Montero:2022vva,
    author = "Montero, Miguel and Parra de Freitas, Hector",
    title = "{New supersymmetric string theories from discrete theta angles}",
    eprint = "2209.03361",
    archivePrefix = "arXiv",
    primaryClass = "hep-th",
    doi = "10.1007/JHEP01(2023)091",
    journal = "JHEP",
    volume = "01",
    pages = "091",
    year = "2023"
}

@article{Minahan:1998vr,
    author = "Minahan, Joseph A. and Nemeschansky, David and Vafa, Cumrun and Warner, Nicholas P.",
    title = "{E-Strings and $\mathcal{N}=4$ Topological Yang-Mills Theories}",
    eprint = "hep-th/9802168",
    archivePrefix = "arXiv",
    journal = "Nucl. Phys. B",
    volume = "527",
    pages = "581--623",
    year = "1998"
}

@article{Polchinski:2003bq,
    author = "Polchinski, Joseph",
    title = "{Monopoles, Duality, and String Theory}",
    eprint = "hep-th/0304042",
    archivePrefix = "arXiv",
    primaryClass = "hep-th",
    doi = "10.1142/S0217751X0401866X",
    journal = "Int. J. Mod. Phys. A",
    volume = "19S1",
    pages = "145--156",
    year = "2004"
}

@article{Banks:2010zn,
    author = "Banks, Tom and Seiberg, Nathan",
    title = "{Symmetries and Strings in Field Theory and Gravity}",
    eprint = "1011.5120",
    archivePrefix = "arXiv",
    primaryClass = "hep-th",
    doi = "10.1103/PhysRevD.83.084019",
    journal = "Phys. Rev. D",
    volume = "83",
    pages = "084019",
    year = "2011"
}

@article{Harlow:2018tng,
    author = "Harlow, Daniel and Ooguri, Hirosi",
    title = "{Symmetries in Quantum Field Theory and Quantum Gravity}",
    eprint = "1810.05338",
    archivePrefix = "arXiv",
    primaryClass = "hep-th",
    doi = "10.1007/s00220-021-04040-y",
    journal = "Commun. Math. Phys.",
    volume = "383",
    pages = "1669--1804",
    year = "2021"
}

\end{document}